\documentclass[12pt,preprint]{aastex}
\usepackage{graphicx}
\title{VLBA Observations of Sub-Parsec Structure in Mrk 231:
Interaction between a Relativistic Jet and a BAL Wind}
\author{Cormac Reynolds\altaffilmark{1}, Brian Punsly\altaffilmark{2}, Preeti Kharb\altaffilmark{3}}
\author{Christopher P. O'Dea\altaffilmark{3} and Joan
Wrobel\altaffilmark{4}}\altaffiltext{1} {Curtin University of
Technology, Department of Imaging and Applied Physics, GPO Box
U1987, Perth, Western Australia, 6102,
Australia}\altaffiltext{2}{4014 Emerald Street No.116, Torrance CA,
USA 90503 and ICRANet, Piazza della Repubblica 10 Pescara 65100,
Italy, brian.punsly@verizon.net or brian.punsly@comdev-usa.com}
\altaffiltext{3}{Department of Physics, Rochester Institute of
Technology, 54 Lomb Memorial Drive,Rochester, NY
14623}\altaffiltext{4}{NRAO, Socorro, NM 87801-0387 United States}
\begin{document}
\begin{abstract} We report on the first high frequency VLBI observations
of the nearby broad absorption line quasar (BALQSO), Mrk 231. Three
epochs of observations were achieved at 15 GHz and 22 GHz, two of
these included 43 GHz observations as well. The nuclear radio source
is resolved as a compact double. The core component experienced a
strong flare in which the flux density at 22 GHz increased by $>
150\%$ (45 mJy) in three months. Theoretical models of the flare
imply that the emission is likely enhanced by very strong Doppler
boosting of a highly relativistic ejecta with a kinetic energy flux,
$Q \sim 3 \times 10^{43} \mathrm{ergs/sec}$. Combining our data with
two previous epochs of 15 GHz data, shows marginal evidence for the
slow advance of the secondary component (located $\approx 0.97$ pc
from the core) over a 9.4 year span. We estimate that the long term
time averaged kinetic energy flux of the secondary at
$\overline{Q}\approx 10^{42}\mathrm{ergs/sec}$. Low frequency VLBA
observations indicate that the secondary is seen through a shroud of
free-free absorbing gas with an emission measure of $\approx 10^{8}
\mathrm{cm}^{-6}\mathrm{pc}$. The steep spectrum secondary component
appears to be a compact radio lobe that is associated with a working
surface between the ram-pressure confined jet, and a dense medium
that is likely to be the source of the free-free absorption. The
properties of the dense gas are consistent with the temperatures,
displacement from the nucleus and the column density of total
hydrogen commonly associated with the BAL wind.
\end{abstract}
\keywords{quasars: absorption lines --- galaxies: active --- accretion disks --- black holes}
\section{Introduction} At a redshift of 0.042, Mrk 231 is one of the nearest radio-quiet quasars
to earth. Mrk 231 is a well studied object because it is one of the
brightest AGN in the infrared sky. It is a quasar that is partially
obscured by dusty gas as evidenced by its red optical/UV spectrum
\citep{lip94,smi95}. The host galaxy has a strong starburst
contribution, but the obscured AGN bolometric luminosity, $L_{bol}$,
can be inferred from a pronounced 10 $\mu$ peak in the IR spectrum
\citep{far03,lon03}. Using the extracted IR AGN luminosity, we
estimate from QSO composite spectra (see \citet{tin05,tel02,zhe97})
that the total thermal luminosity created by the central engine is
$L_{bol} \lesssim 2\times 10^{46} \mathrm{ergs/s}$ and Mrk 231 would
appear as a quasar with $M_{V} \approx -25.3$ if not for the
semi-transparent dusty shroud (in agreement with the estimate in
\citet{bok77})\footnote{Throughout this paper, we adopt the
following cosmological parameters: $H_{0}$=71 km/s/Mpc,
$\Omega_{\Lambda}=0.73$ and $\Omega_{m}=0.27$ in all our
calculations. At z = 0.042, 1 mas angular separation equates to
0.818 pc displacement in the sky plane in the adopted cosmology.}.
This equates to about 2.7 magnitudes of visual extinction created by
the dusty enveloping gas. Comparing the composite spectrum to the
optical/UV spectrum in \citet{smi95} indicates even more severe
reddening in the blue band, 3.1 magnitudes at $4400 \AA$. It is the
closest radio quiet quasar in which there is sufficient radio flux
to permit high signal to noise imaging with high frequency VLBI
(Very Long Baseline Interferometry) observations. As a result of the
ability to utilize the high resolution of VLBI, this nucleus can be
resolved with higher precision than any other radio quiet quasar
nucleus.\footnote{The radio loudness, $R$, is usually defined as a 5
GHz flux density 10 times larger than the $4400 \AA$ flux density,
$R=S_{5 \mathrm{GHz}}/S_{4400 \AA}<10$ \citep{kel89}. This
definition is not appropriate, if there is intrinsic extinction as
in BALQSOs. One must correct for the intrinsic blue band extinction
and recompute the $R$ with dereddened blue fluxes, \citep{bec00}.
Noting the blue band extinction in the text and using the historic 5
GHz flux densities from NED yields a dereddened radio loudness, $1.4
< R < 3.8$. This is close to the radio intermediate quasar category,
$10 < R < 250$, which are often just Doppler boosted radio quiet
quasars \citep{fal96}. We show that Doppler boosting greatly
enhances $R$ in Mrk 231 as well. The intrinsic radio quiet nature of
the AGN in Mrk 231 is not directly apparent because of optical
reddening and Doppler boosting.} Our 43 GHz observations can resolve
the nucleus to within $5\times 10^{17}\mathrm{cm}$ ($\sim$ 0.16 pc).
It is a fortuitous circumstance that this object is a broad
absorption line quasar (BALQSO) as well. About 10\% - 15\% of
quasars show broad absorption line (loosely defined as absorbing gas
that is blue shifted at least 5,000 km/s relative to the QSO rest
frame and displaying a spread in velocity of at least 2,000 km/s)
outflows \citep{wey97}. It is widely believed that most or all radio
quiet quasars have BAL outflows, but the designation of a quasar as
a BALQSO depends on whether the line of sight intersects the solid
angle subtended by the BAL outflow. Mrk 231 is a LoBALQSO (a BALQSOs
that exhibits low ionization resonant absorption troughs). Mrk 231
has broad absorption in a variety of low ionization species, Mg II
$\lambda$ 2798, the well known Na I D doublet, He I $\lambda$3889 I,
Ca II K $\lambda$3934 and Ca II H $\lambda$3968
\citep{smi95,lip94,mar97}. Thus, Mrk 231 displays the full panoply
of quasar related phenomena, broad emission lines, powerful thermal
luminosity, UV broad absorption lines and a radio jet. Therefore, as
a consequence of its proximity to earth, Mrk 231 can provide an
optimal opportunity to study the interaction of these physical
elements by means of high resolution VLBI observations on sub-parsec
scales.
\par Earlier 15 GHz self-calibrated VLBA (Very Long Baseline
Array)\footnote{The National Radio Astronomy Observatory is a
facility of the National Science Foundation operated under
cooperative agreement by Associated Universities, Inc.} observations
reported in \citet{ulv99,ulv00} with sub mas resolution revealed a
double radio source separated by approximately 1 pc. We initiated a
program to look for more sub-structure and component motion at
higher resolution by means of phase-referenced observations at 22
GHz and 43 GHz VLBA as discussed in section 2. No further resolution
into sub-components was achieved even at 43 GHz. The two archival
epochs showed that the nucleus was very dynamic and set a lower
bound on the time variability brightness temperature of $T_{b} >
10^{10}\,\mathrm{K}$, below the inverse Compton limit, $\sim
10^{12}\,\mathrm{K}$ \citep{ulv99,ulv00}. Thusly motivated, our
program of observations attempted short time interval sampling (3
months between the epoch 2006.07 and 2006.32) in order to elucidate
this highly dynamic behavior. Mrk 231 did not disappoint, there was
an enormous flare between these two epochs. Section 5 is dedicated
to analyzing the rapidly changing core. We use our data and the
adopted cosmological model to show that the strong flare will
violate the constraints associated with the brightness temperature
limit unless relativistic Doppler beaming is invoked.
\par We also explore the steep spectrum secondary component of the nucleus of
Mrk 231 in section 6. It is mildly variable and it is just less than
a parsec from the core. Based on the nature of the relativistic
core, it is natural to wonder if there is any relativistic motion of
the secondary. In section 3, we determine that the secondary is
either not moving at all or is moving at velocities far below
relativistic values. Thus, the more relevant question becomes what
is constraining the secondary from having such motion. Section 6
uses our VLBA observations and archival low frequency observations
to model the apparent free-free absorption of the secondary
emission. We argue that this same gas is confining the secondary by
ram pressure and that this gas is likely one and the same as the BAL
wind.
\section{The Observations}
High frequency (15, 22 and 43~GHz) observations were conducted at
epochs 2006.07 (Project ID BP124A) and 2006.32 (Project ID BP124B)
with all ten antennas of the VLBA. In addition, data from 2000.02
(Project ID BU013E) at frequencies of 8.4, 15 and 22~GHz and 8.3 GHz
data from 2006.67 (Project ID BA080B) were retrieved from the VLBA
public archive. The data were correlated on the VLBA correlator in
Socorro, NM and calibrated with NRAO's Astronomical Image Processing
System (AIPS) software package. In addition to the standard
calibration procedure for VLBI polarization observations
\citep{cotton1993, leppanen1995}, an attempt was made to correct for
amplitude losses due to opacity using the AIPS task APCAL
\citep{leppanen1993}.

At all epochs and frequencies, except 2006.67, the data were initially
phase-referenced \citep{beasley1995} to J1302+5748. At frequencies below 43~GHz
the data were then phase self-calibrated, but at 43~GHz there was insufficient
signal-to-noise to allow this. The phase reference source was imaged to provide
the best possible model for the phase referencing procedure, but proved in any
case to be a very close approximation to a point source.

\begin{figure}
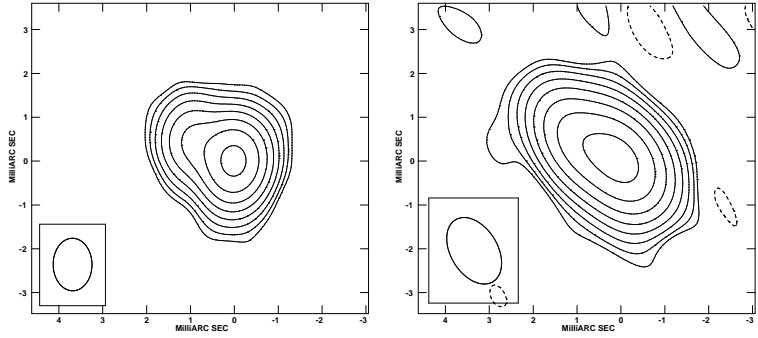

\includegraphics[height=50mm,angle=-90,keepaspectratio]{./bu013e_8ghz.ps}
\includegraphics[height=50mm,angle=-90,keepaspectratio]{./ba80_8ghz.ps}
\caption{VLBA 8~GHz images of Mrk~231 from epoch 2000.02 (left) and
epoch 2006.67 (right). The contours start at 0.7 mJy/beam and
increase in factors of 2 to a maximum of 89.6 mJy/beam.}
\label{8ghzmaps}
\end{figure}

\begin{figure}
\includegraphics[height=53mm,angle=-90,keepaspectratio]{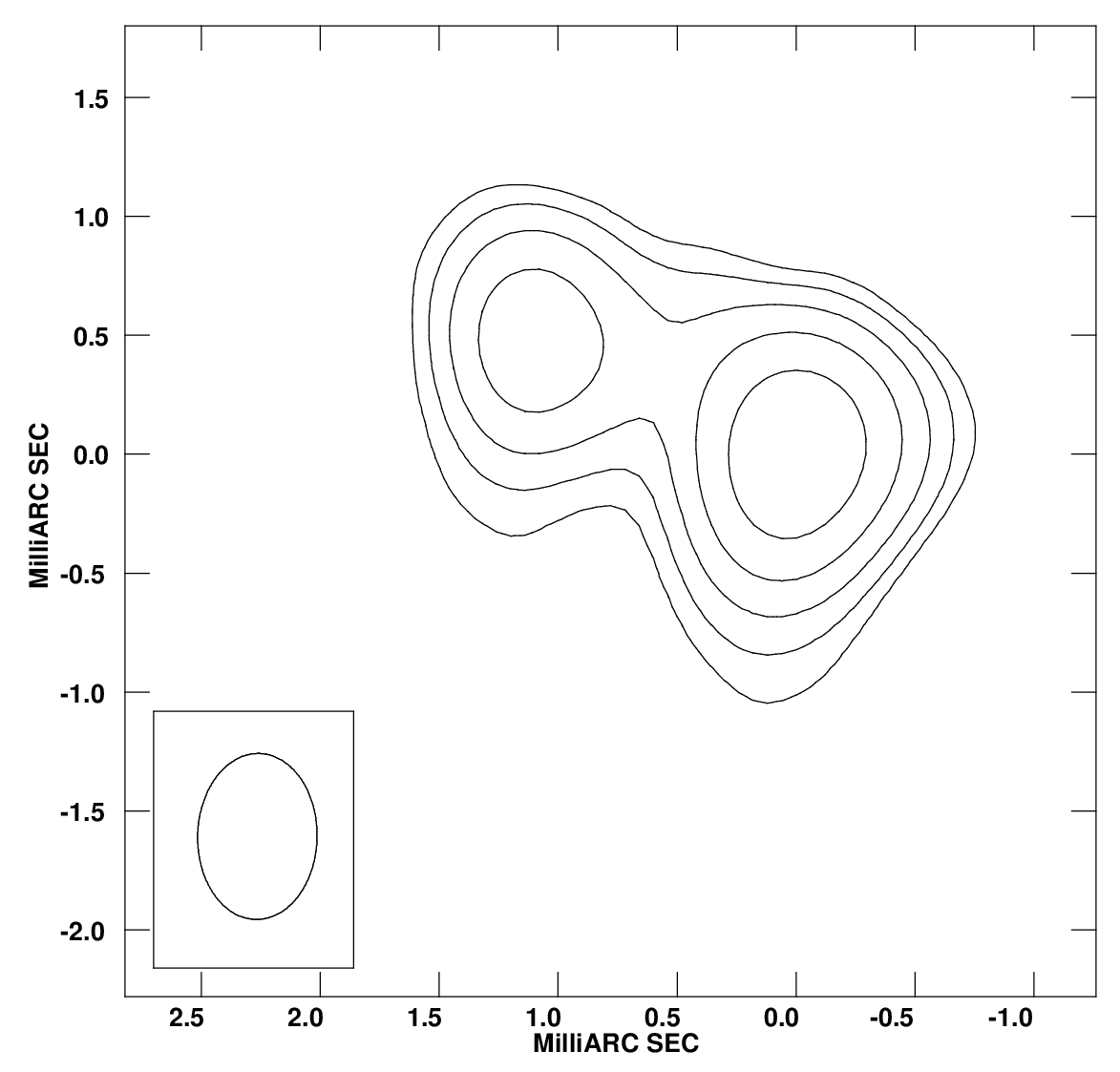}
\includegraphics[height=53mm,angle=-90,keepaspectratio]{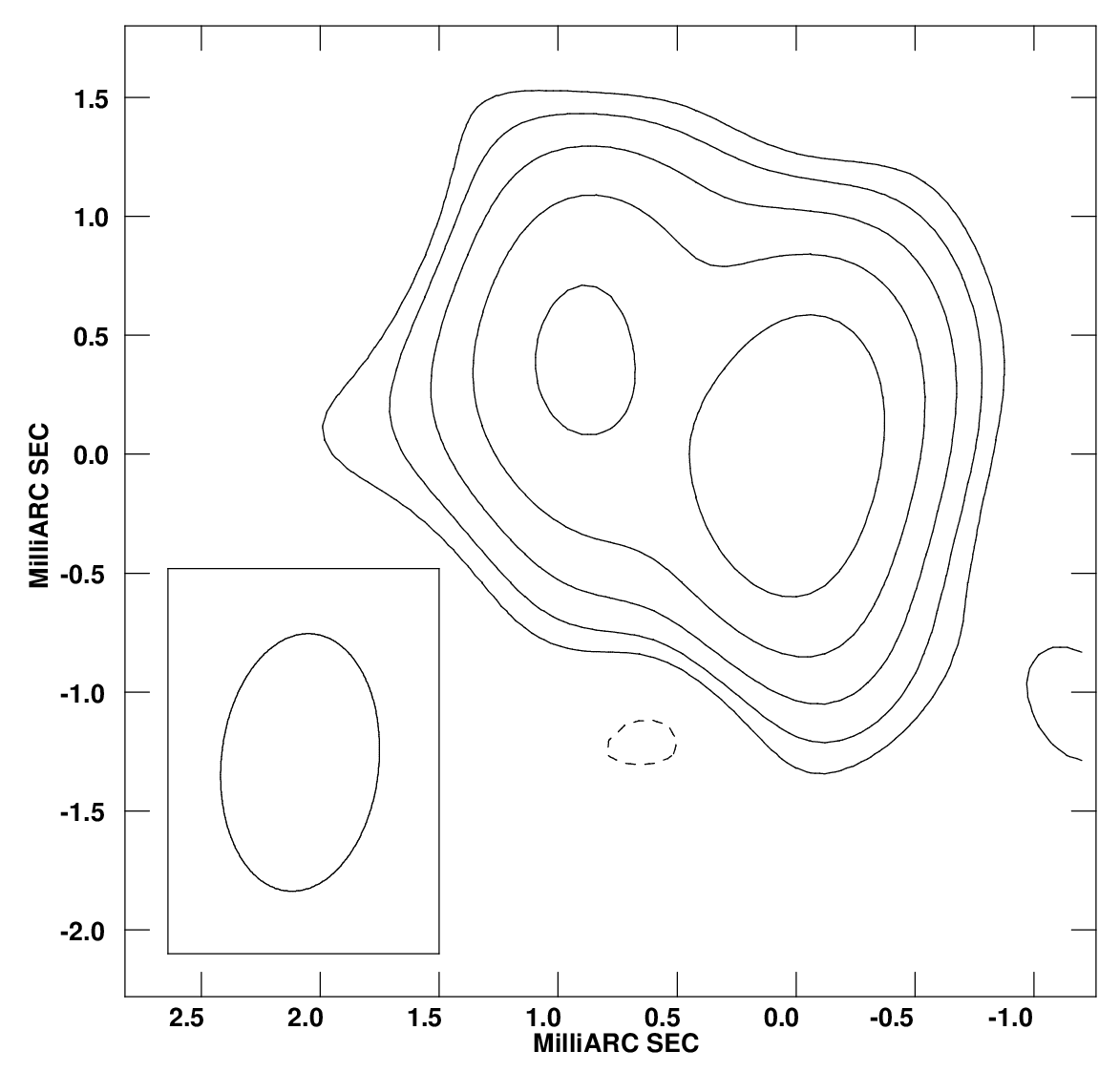}
\includegraphics[height=53mm,angle=-90,keepaspectratio]{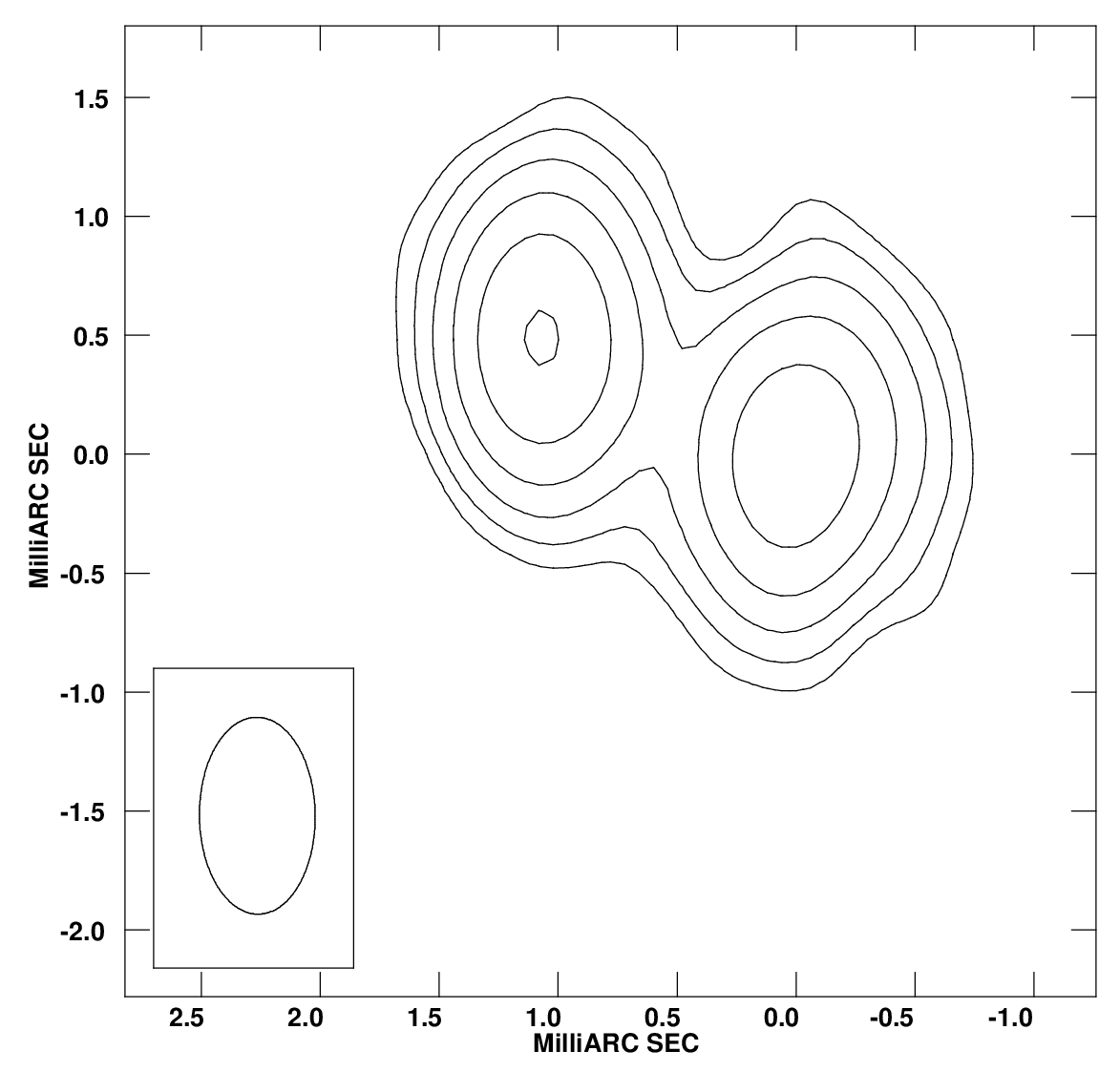}
\caption{VLBA 15~GHz images of Mrk\,231 from epoch 2000.02 (left),
epoch 2006.07 (middle) and epoch 2006.32 (right). The contours start
at 1.8 mJy/beam and increase in factors of 2 to a maximum of 57.6
mJy/beam.
\label{15ghzmaps}
}
\end{figure}

Figure \ref{8ghzmaps} captures the images of the 8.4~GHz observations from
2000.02 and 2006.67. Figures~\ref{15ghzmaps} and \ref{22ghzmaps} are the images
at 15 and 22~GHz, respectively, for all three epochs and
Figure~\ref{43ghzmaps} is the 43~GHz images from 2006.07 and
2006.32. All images of Mrk\,231 presented in this paper are made
with Briggs' robust weighting scheme \citep{bri95}, with the
robustness parameter set to zero, with the exception of the 43~GHz
image at epoch 2006.32 where natural weighting is used. All images
are centered on the position of the southwestern, secondary,
component at rectangular coordinate (0~mas, 0~mas).

Each of the images shows a pair of unresolved components separated by
about 1~mas that vary in brightness, but do not vary significantly in
separation. The NE component is the most variable and has the flattest spectrum
and it is assumed to be associated with the AGN nucleus.

Gaussians were fitted to the positions of the two components using
the DIFMAP software package \citep{shepherd94, shepherd95} and the
resultant parameters are presented in Tables~\ref{table:fluxes} and
\ref{table:sep}. Errors in the flux density in Table 1 arise from
two sources: the absolute flux scale error, and an error due to
uncertainties in the model fit. The method used to determine the
model fit part of the error is from \citet{bri07}. The absolute flux
scale errors for the VLBA are estimated to be $\sim 5\%$ at 8~GHz
\citep{ulv08} and 15~GHz \citep{hom02}, and $\sim 7\%$ at 22~GHz
\citep{hom02}. At 43~GHz we estimate the error on the absolute flux
density scale to $\sim -8\%/+15\%$ -- the non-symmetric error
arising from the fact that some flux density is likely lost due to
the inability to phase self-calibrate this source at that frequency. An
analysis of the amplitude corrections derived from self-calibration of various
calibrator sources indicates that these observations were not affected by any
unusual amplitude errors that would lead to larger errors than the values
quoted above.

\begin{figure}
\includegraphics[height=53mm,angle=-90,keepaspectratio]{./bu013e_22ghz.ps}
\includegraphics[height=53mm,angle=-90,keepaspectratio]{./bp124a_22ghz.ps}
\includegraphics[height=53mm,angle=-90,keepaspectratio]{./bp124b_22ghz.ps}
\caption{VLBA 22~GHz images of Mrk\,231 from epoch 2000.02 (left),
epoch 2006.07 (middle) and epoch 2006.32 (right). The contours start
at 1.8 mJy/beam and increase in factors of 2 to a maximum of 57.6
mJy/beam.
\label{22ghzmaps}
}
\end{figure}

\begin{figure}
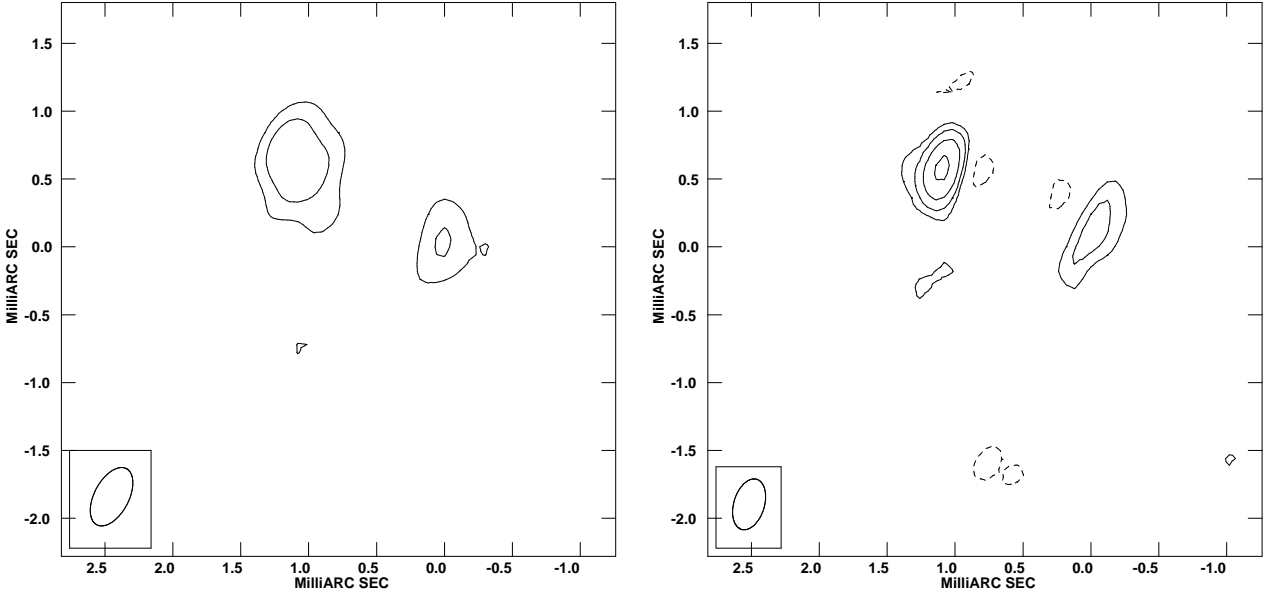

\includegraphics[height=85mm,angle=-90,keepaspectratio]{./bp124a_43ghz.ps}
\includegraphics[height=85mm,angle=-90,keepaspectratio]{./bp124b_43ghz.ps}
\caption{VLBA 43~GHz images of Mrk\,231 from epoch 2006.07 (left) and
epoch 2006.32 (right). The contours start at 2.2 mJy/beam and
increase in factors of 2 to a maximum of 17.6 mJy/beam.
\label{43ghzmaps}
}
\end{figure}

\begin{table}
\caption{High Frequency VLBA Observations Data Summary \label{table:fluxes}}\tiny{
{\begin{tabular}{cccccc} \tableline \rule{0mm}{3mm}
 Component &  Date & 8.4 GHz Flux & 15 GHz Flux & 22 GHz Flux & 43 GHz Flux  \\
 &    &  Density (mJy)/\% P\tablenotemark{a}& Density (mJy)/\% P\tablenotemark{a}& Density (mJy)/\% P\tablenotemark{a}& Density (mJy)/\% P\tablenotemark{a} \\
\tableline \rule{0mm}{3mm}
 NE & 1996.93 & $18 \pm 2$ & $17 \pm 2$ & ...  & ...  \\
 NE & 1998.71 & ... & $44 \pm 3$  &  ...   & ...  \\
 NE & 2000.02 & $29.1 \pm 1.5 / < 1.0$ & $ 31.3 \pm 1.8/ < 1.9$ & $26.7 \pm 2.0 \ < 3.9$ &  ...  \\
 NE & 2006.07 & ... & $43.1 \pm 2.2/< 1.9$ & $29.4 \pm 2.2/< 3.4$ & $20.1^{+3.2}_{-1.9}/ < 13.7$  \\
 NE & 2006.32 & ... & $68.6 \pm 3.5/ < 1.2$ & $74.8 \pm 5.3 / < 0.7$ & $21.3^{+3.3}_{-1.8}/ <10.1$ \\
 NE & 2006.67 & $57.0 \pm 3.9$ /... & ... & ... & ... \\
 \tableline \rule{0mm}{3mm}
 SW & 1996.93 & $114 \pm 11$ & $51 \pm 3$ & ...  & ...  \\
 SW & 1998.71 & ... & $60 \pm 3$  &  ...   &.... \\
 SW & 2000.02 & $126.2 \pm 6.3 / <0.24 $ & $ 71.2 \pm 3.7\ <0.28 $ & $36.5 \pm 2.7 \ < 2.9$ &.... \\
 SW & 2006.07 & ... & $69.5 \pm 3.5/< 1.2$ & $43.1 \pm 3.1 /< 1.3 $ & $9.8^{+1.8}_{-1.3}/ <28.1$  \\
 SW & 2006.32 & ... & $66.8 \pm 3.4 /< 1.3$ & $40.0 \pm 2.9/ <9$ & $8.6 ^{+1.4}_{-0.9}/ < 25.1$ \\
 SW & 2006.67 & $127.5 \pm 6.3$ /... & ... & ... & ... \\
 \tableline \rule{0mm}{3mm}
\end{tabular}}
\tablenotetext{a}{The polarization is a 3 $\sigma$ upper limit}}
\end{table}

\begin{table}
\caption{Component Separation \label{table:sep}}{\tiny
{\begin{tabular}{ccccccccccccc} \tableline \rule{0mm}{3mm}
 Date & 8.4 GHz  \tablenotemark{a}   &    $\sigma_{1}$ \tablenotemark{b}  & $\sigma_{2}$\tablenotemark{c}   & 15 GHz  \tablenotemark{a} &    $\sigma_{1}$\tablenotemark{b}  & $\sigma_{2}$\tablenotemark{c}   & 22 GHz \tablenotemark{a}    &    $\sigma_{1}$\tablenotemark{b}  & $\sigma_{2}$\tablenotemark{c}   & 43 GHz \tablenotemark{a}    &    $\sigma_{1}$\tablenotemark{b}  & $\sigma_{2}$\tablenotemark{c}     \\
\tableline \rule{0mm}{3mm}

 1996.93 & ... & ... & ...   & 1.081 & 0.03 & 0.172 & ... & ... & ...   & ... & ...  & ... \\
 1998.71 & ... & ...  & ... & 1.162 & 0.004 & 0.174   &  ... & ... & ...  & ... & ... & ...   \\
 2000.02 & 1.038 & 0.004 & 0.126 & 1.173 & 0.003 & 0.139 & 1.219  &0.005 & 0.128 &  ... & ...   & ... \\
 2006.07 & ... & ... & ... & 1.009 & 0.006 & 0.172 & 1.093 & 0.005 &0.132 & 1.222 & 0.013 & 0.142  \\
 2006.32 & ... & ... & ... & 1.172 & 0.002 & 0.089 & 1.193 & 0.002 & 0.059 & 1.216 & 0.010 & 0.103  \\
 2006.67 & 1.124 & 0.016 & 0.267 & ... & ... & ... & ... & ... & ...   & ... & ... & ...  \\

\end{tabular}}
\tablenotetext{a}{Measured separation at the designated frequency in
mas}

\tablenotetext{b}{$\sigma_{1}$ is the error computed using the
technique of \citet{con98}, see equation (1)}
 \tablenotetext{c}{$\sigma_{2}$ is the error computed using the technique of \citet{lob05}, see equation (2)}}
\end{table}

\par Based on the inverted spectrum from epoch 2000.02, we anticipated dramatic
behavior on the smallest of scales. Thus, we proposed for two epochs
of high frequency observations that were closely spaced in time,
2006.07 and 2006.32. The observations were phase referenced to
J1302+5748 at 15, 22 and 43 GHz. We also gathered polarization data.
There is no evidence that the flux density recovered at any of our
frequencies was strongly dependent on the weighting used, indicating
that we are dealing with two compact components with little, if any,
diffuse flux on the range of spatial scales probed by these
observations.

The inability to phase self-calibrate at 43~GHz opens the possibility that some
flux density may have been lost due to phase errors. In lieu of the
undetected secondary calibrator, J1306+5529, one can use the
secondary component to the SW in MRK\,231 as a crude amplitude
calibrator. Note the flux density of the secondary was the same
between epochs 2000.02 and 2006.32 at 22~GHz to within a few percent
in Table 1 indicating that it is not strongly variable. Therefore, we treat
this component as approximately steady at high frequency and it can potentially
serve as a crude calibrator for the core flux density at 43 GHz. The flux
density of the secondary (our crude high frequency amplitude calibrator)
measured at both epochs at 43~GHz agrees to within 1.2~mJy (0.6 sigma). This
supports the suggestion that the phase referencing with respect to J1302+5748
recovered the majority of the flux at 43~GHz, and any amplitude loss due to
phase errors is likely to be similar at the two epochs.
\par We note that in epoch 2006.07 bad weather and technical problems allowed
us to recover data from effectively just 5.25 of the 10 array
antennas at 43~GHz, so these data have sparser u-v coverage and an
increased RMS noise level. Our 43~GHz images in
Figure~\ref{43ghzmaps} show partial resolution of the core at epoch
2006.07 with an extension at PA = $135\,^{\circ}$ from NE to SW.
Unfortunately, we must consider this detection unreliable due to the
poor beam shape and the low SNR. We stress that the data from all 10
antennas were recovered at 43 GHz at epoch 2006.32. This higher
quality data is used extensively in section 5. \par In Figures 5 and
6, we plot the spectrum of the core and the secondary, respectively,
for all three epochs of observation as well as the two archival data
sets in \citet{ulv99,ulv00}.

\begin{figure}
\includegraphics[width=120 mm, angle= -90]{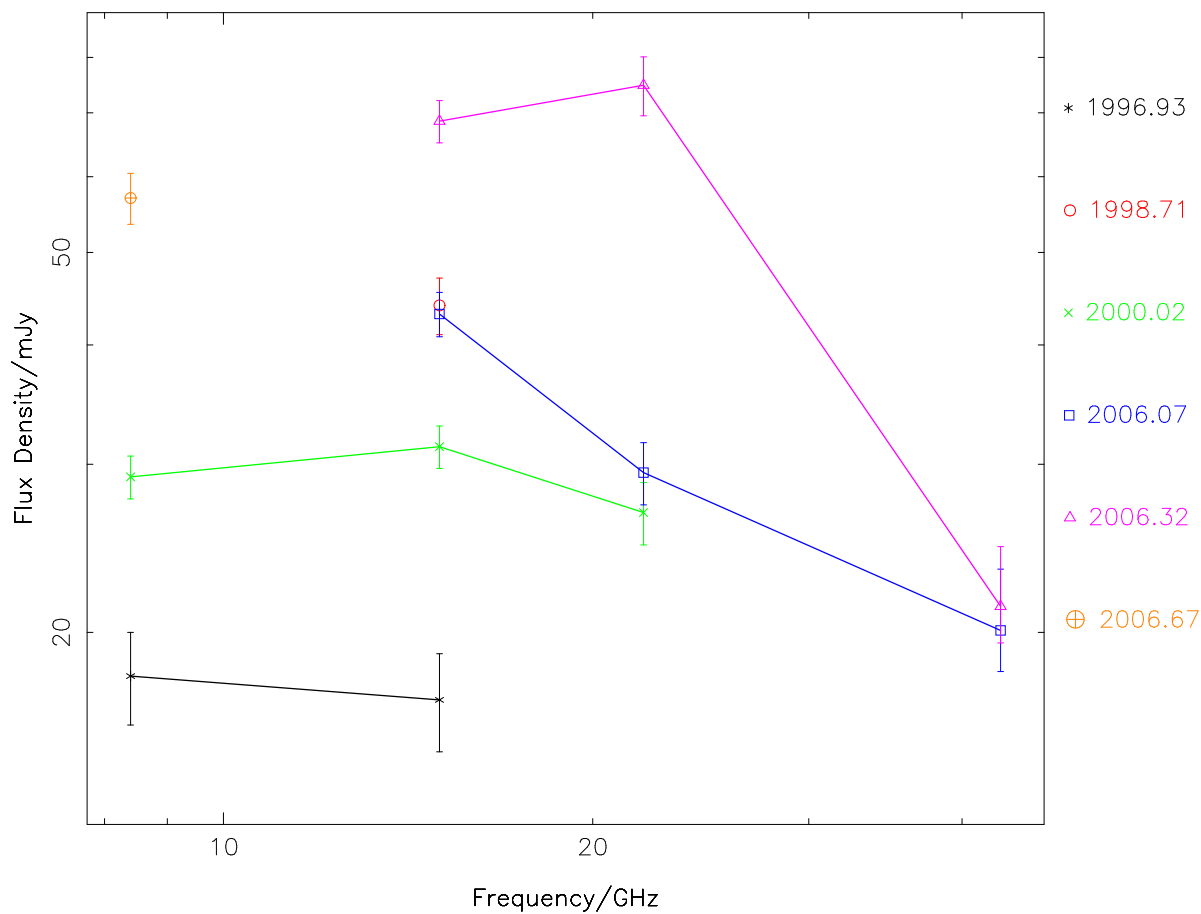}
\caption{The core spectra of Mrk 231 at six different epochs of VLBA
observation. Notice that the data point for 1998.71 overlaps the 15
GHz point at epoch 2006.07.}
\end{figure}
The core spectra in Figure 5 are highly variable in magnitude. All
the epochs reveal an optically thick core that is indicated by flat
or inverted spectra in a finite band of frequency. Clearly the most
powerful epoch is 2006.32 which we explore in detail in section 5.
In contrast Figure 6 shows that the secondary is optically thin,
steep spectrum at all epochs. The data are supportive of a secondary
that is mildly variable, characterized by changes of $< 10\%$ ($<
1\sigma$) on time scales of 0.25 years, $\sim 15\%$ ($< 2\sigma$) on
time scales of 6 years and $\sim 30\%$ on time scales of 9.4 years.
Notice that the spectra show significant curvature, reflecting a
deficit of high frequency flux relative to a power law extrapolated
from lower frequency.
\begin{figure}
\includegraphics[width=120 mm, angle= -90]{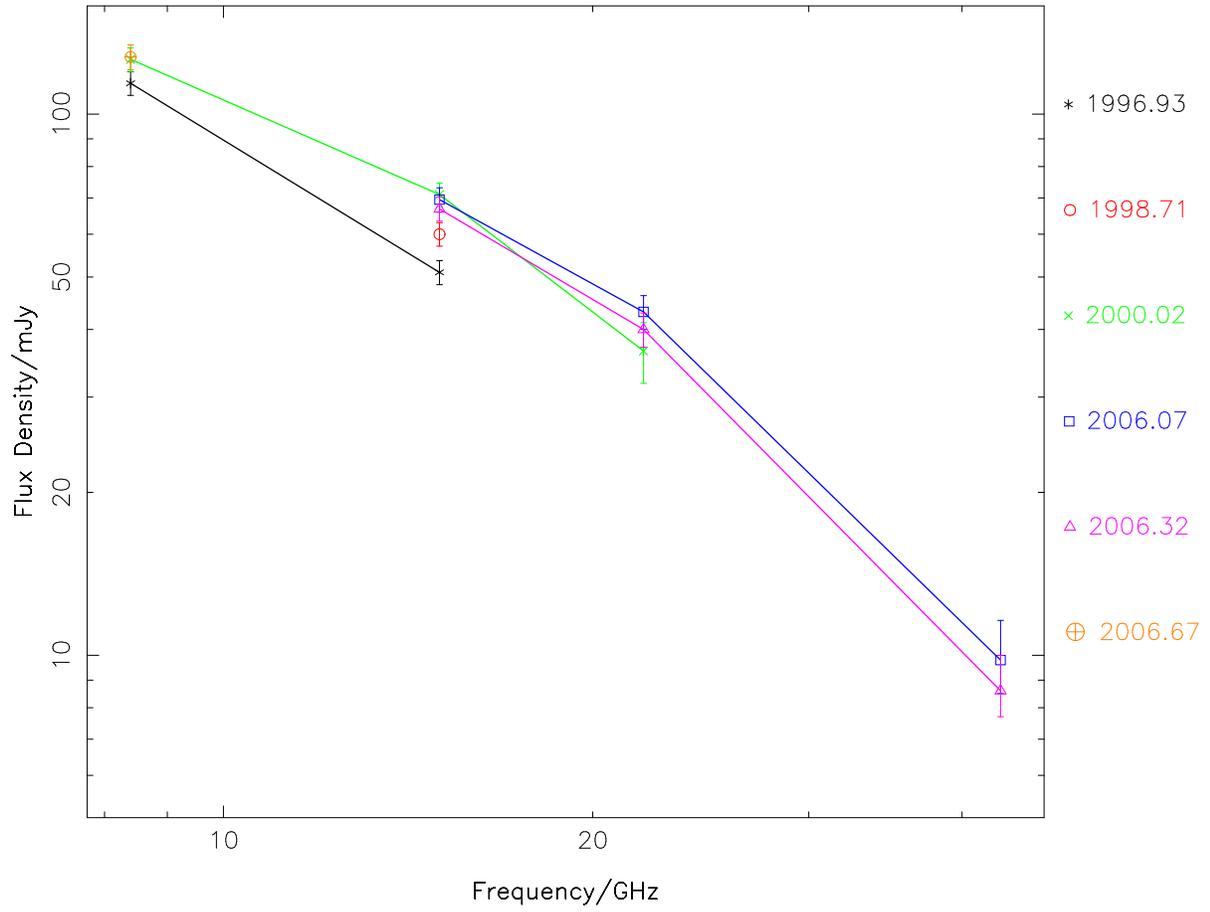}
\caption{The spectra of the secondary component of the nucleus of
Mrk 231 at six different epochs of VLBA observation.}
\end{figure}

\section{Component Motions and Positions} An important part of this analysis is to try to
determine the motion of the secondary relative to the core over the
9.4 year history of VLBA observations that were capable of resolving
the compact nuclear double (i.e., observations at $\geq$ 15 GHz).
The entries in Tables 1 and 2 include the historical data from
\citet{ulv99,ulv00} for this reason. It is clear that there is very
little component motion. Therefore, a careful analysis of the errors
that arise in the determination of the component separation is
critical. The displacements between the core and secondary are
entered in Table 2. Those entries in the first five rows have errors
computed by a method that is based on the error in discerning the
position of the centroid of an elliptical Gaussian fit to the data
in the presence of background noise (e.g., \citet{con98}),
\begin{eqnarray}
&&\Delta\theta_{\mathrm{sep}}=0.5\left[\left(\frac{\theta_{\mathrm{beam}}}
{SNR_{\mathrm{core}}}\right)^{2}+\left(\frac{\theta_{\mathrm{beam}}}{SNR_{\mathrm{secondary}}}\right)^{2}\right]^{0.5}\;,
\end{eqnarray}
where $\Delta\theta_{\mathrm{sep}}$ is the error in the estimate of
the separation in mas, $\theta_{\mathrm{beam}}$ is the beamwidth in
mas and SNR is the signal to noise ratio. Notice that the component
separations in Table 2 widen with increasing observing frequency at
each epoch as expected for an optically thick core
\citep{lob98,kon81}. Thus, in computing a change in the component
separation between epochs a common frequency should be used. The
lone common frequency of observation for all epochs is 15 GHz.
\par However, the method described by equation (1)
does not account for core subcomponents that flare on scales below
the resolution limit of the beam without any actual motion
occurring. If a subcomponent flares, the centroid of the elliptical
fit of the core can shift in position and give the illusion of
secondary motion. This is a concern for Mrk 231 since Figure clearly
indicates the presence of strong core flares that are not resolved
by VLBI. For example, the core flux density flares at 1998.71
relative to the flux density of the core at 1996.93 and the
components appear to separate by 0.08 mas at 15 GHz according to
Table 2. This can be described by the base of the jet at the far
eastern edge of the core being the source of the flare with no
actual secondary motion. The position of the elliptical fit just
shifts to the east by about 0.08 mas, well below the resolution of
the beam at 15 GHz or 22 GHz where the flare is detected. Thus, we
need an estimate of the error in component separation that
incorporates the maximum size subcomponent flare in the core that
can not be resolved with VLBI (equivalently, the minimum size
subcomponent that can be resolved by VLBI). Comparing the maps of
the compact double in \citet{ulv99} at 15.3 GHz (resolved), 8.4 GHz
(partially resolved) and 5 GHz (unresolved), it appears that the
uncertainty posed by unresolved subcomponents in the core will be
about 1/2 - 1/3 of the beamwidth projected along the direction
toward the secondary. Fortuitously for the compact double in Mrk 231
this direction is close to the minor axis of the beam. In general,
the minimum size subcomponent that is resolvable by VLBI will depend
on the SNR, so we look for a more general and quantitative
formulation. To this end we invoke the technique of \citet{lob05}
that considers the ability of an interferometer to resolve
components modeled by elliptical fits in presence of noise assuming
a rectangular power law interferometric visibility sampling
function,
\begin{eqnarray}
&&\theta_{\mathrm{lim}}= \frac{2^{2-\beta/2}}{\pi}\left[\pi a b
\ln{2} \ln{\left(\frac{SNR}{SNR-1}\right)}\right]^{0.5}\;.
\end{eqnarray}
In equation (2), $\theta_{\mathrm{lim}}$ is the minimum size
subcomponent in mas that can be resolved by the interferometer. In
our calculations we have the weighting of the data, $\beta=0$
corresponding to unweighted data. The major axis and minor axis of
the beam in mas are $a$ and $b$, respectively. The values of
$\theta_{\mathrm{lim}}$ that we derive from equation (2) range from
$b/4$ - $b/2$ for the observations in Table 2. We consider,
$\theta_{\mathrm{lim}}$ an estimate of the uncertainty in the core
position due to unresolved flaring subcomponents. The secondary does
not flare, so we do not consider this a likely source of error in
the position of the secondary. We add this error to the error
introduced in equation (1) in quadrature ($\theta_{\mathrm{lim}}$ is
the dominant contributor). The results computed by the formula, (2),
are listed for the separation error on rows 6 - 10 of Table 2. They
are large compared to the variations in the component separation
between any pair of epochs. We conclude that the observations in
Table 2 are not precise enough to determine the rate of component
separation, except as an upper limit determined by
$\theta_{\mathrm{lim}}$. The VLBA observations indicate that the
separation in components changed in 9.4 years (from 1996.93 to
2006.32) by
\begin{eqnarray}
&& \Delta\theta = 0.091 \pm 0.094\; \mathrm{mas}\,.
\end{eqnarray}
In arriving at the result given by equation (3), we noted that the
core at epoch 1996.93 was at an extreme historic low. Thus, we do
not consider core flares relevant in this quiescent state.
Therefore, we use the error derived from equation (1) that appears
in the first row of Table 2 as an accurate assessment of the error
in the measured component separation at this epoch. At 2006.32,
there is clearly a flare, so we use the 15 GHz entry derived from
equation (2) that appears in row 10 of Table 2 to estimate the error
in the component separation at this epoch. The secondary separation
in equation (3) equates to a time averaged secondary separation
velocity, $v_{adv}$
\begin{eqnarray}
&& v_{adv} = 0.026c \pm 0.027c\;.
\end{eqnarray}
Our data is consistent with no apparent motion of the secondary relative to the core. It should
be remembered that $v_{adv}$ is the projection of the separation velocity of the
components onto the sky plane.
\section{The Core Structure at 2006.07} Table 2 indicates that the
component separation increases with observing frequency at all
epochs, This is expected in optically thick cores as the $\tau =1$
surface for synchrotron self absorption moves outward relative to
the true core position with decreasing frequency \citep{kon81}. In
order to ascribe a quantitative measure of this phenomena one must
resort to highly model dependent estimates that involve much
speculation about many unknown physical quantities that parameterize
the putative jet \citep{lob98}. However, notice that this variation
is particularly pronounced at 2006.07. Furthermore, from Figure 5,
the core is actually optically thin between 15 GHz and 22 GHz, yet
there is an 84 $\mu$as differential in the component separations. In
this section we study the ramifications of this to see if we can
make some quantitative statements with a minimal amount of model
dependent assumptions.
\par We begin by overlaying the component positions on the 43 GHz
radio image at epoch 2006.07 in Figure 7.
\begin{figure}
\includegraphics[width=150 mm, angle= 0]{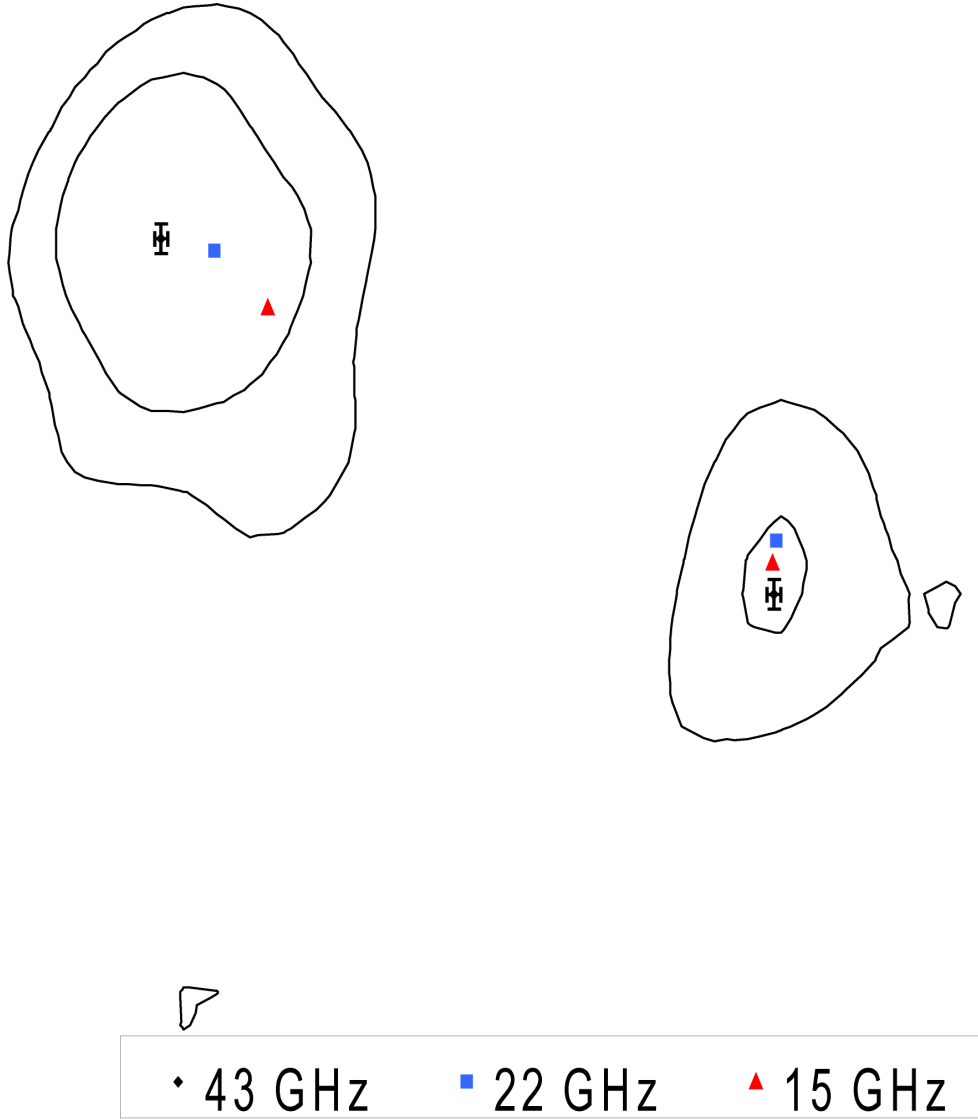}
\caption{An overlay of the frequency dependent component positions
on the 43 GHz image at epoch 2006.07. The errors in positions are
naively computed using equation (1). Only the 43 GHz points have
significant error bars. A more accurate error assessment method is
shown in Figure 8 and discussed in the text.}
\end{figure}
The main issue is how does one register the image centers at
different observing frequencies. This was not designed as an
astrometric experiment and we do not have a good handle on the
positional jitter that is induced by fluctuations in the
tropospheric delay.
\par Ostensibly there is no obvious way to place
errors on the component positions in Figure 7. However, notice that
the secondary components are fairly close in position. In fact we
can argue that they should be even closer than this based on the
steep spectrum in Figure 6. We seek to register all of the epoch
2006.07 radio images relative to the 22 GHz position of the
unresolved secondary at 2006.07. Let $\Sigma_{22}$ be the region
containing the plasma that emits the bulk of the 22 GHz secondary
emission. For any reasonable ensemble of energetic particles, there
must also be emission at 43 GHz and 15 GHz. First, consider the 43
GHz emission. The spectral index in Figure 6 between 22 GHz and 43
GHz is incredibly steep, $\alpha_{43}^{22} = 2.2$, where the
spectral index is defined in terms of the flux density by the
convention, $S_{\nu} \sim \nu^{-\alpha}$. Between 15 GHz and 22 GHz
the spectral index is significantly flatter, but still very steep,
$\alpha_{22}^{15} = 1.2$. Thus, unless there is an incredibly abrupt
spectral break above 22 GHz and a second weak 43 GHz component,
virtually all of the 43 GHz radiation from the secondary would be
emitted from the particles in $\Sigma_{22}$. Therefore, we consider
the assumption that the 43 GHz emitting electrons are co-spatial
with the 22 GHz emitting plasma is most likely valid. We can not
make the same argument for the 15 GHz emitting electrons. Since the
spectrum is so steep between 43 and 22 GHz and the spectrum above 22
GHz is also steep, we make the conservative assumption that the 15
GHz flux density from $\Sigma_{22}$ is at least as big as the 22 GHz
flux density, 43.1 mJy. With this conservative assessment, we can
estimate an upper bound on the error in the 15 GHz secondary
position if it is re-registered co-spatially with the 22 GHz
secondary position. Namely, if the remaining 26.4 mJy of residual 15
GHz flux resides primarily in a disjoint component, $\Sigma_{15}$,
this would create the maximum offset in the true absolute position
of the 15 GHz component position relative to $\Sigma_{22}$. The 15
GHz secondary is unresolved in Figure 2, and according to
\citet{ulv05} and our own experience discussed in section 3, if
$\Sigma_{15}$ was half a beam width away from $\Sigma_{22}$, the
secondary would be partially resolved in the radio map at 15 GHz.
There is no partially resolved component in Figure 2. Thus, we can
set an upper bound on the error in registering the 15 GHz secondary
position to the 22 GHz position, by positing that 26.4 mJy of flux
density arise from an optically thin component, $\Sigma_{15}$, that
lies one half of a beam width from $\Sigma_{22}$ (which accounts for
the remaining 43.1 mJy of the 15 GHz flux density of the secondary).
Then taking the flux density weighted average of the component
positions (i.e., offset from $\Sigma_{22}$ =
$(26.4/69.5)(\theta/2))$ yields a conservative upper bound on the
offset between the true 15 GHz position and the 22 GHz position. The
resultant upper bound on the error in the re-registered 15 GHz
position is 0.11 mas and 0.018 mas in the x and y directions,
respectively. Figure 8 is the Figure 7 overlay re-registered to the
22 GHz secondary position (i.e., all three of the secondary
positions agree). The errors on the re-registered 22 and 43 GHz core
data points in Figure 8 arise from equation (1). The error on the 22
GHz core position is too small to be seen in this figure.
\begin{figure}
\includegraphics[width=150 mm, angle= 0]{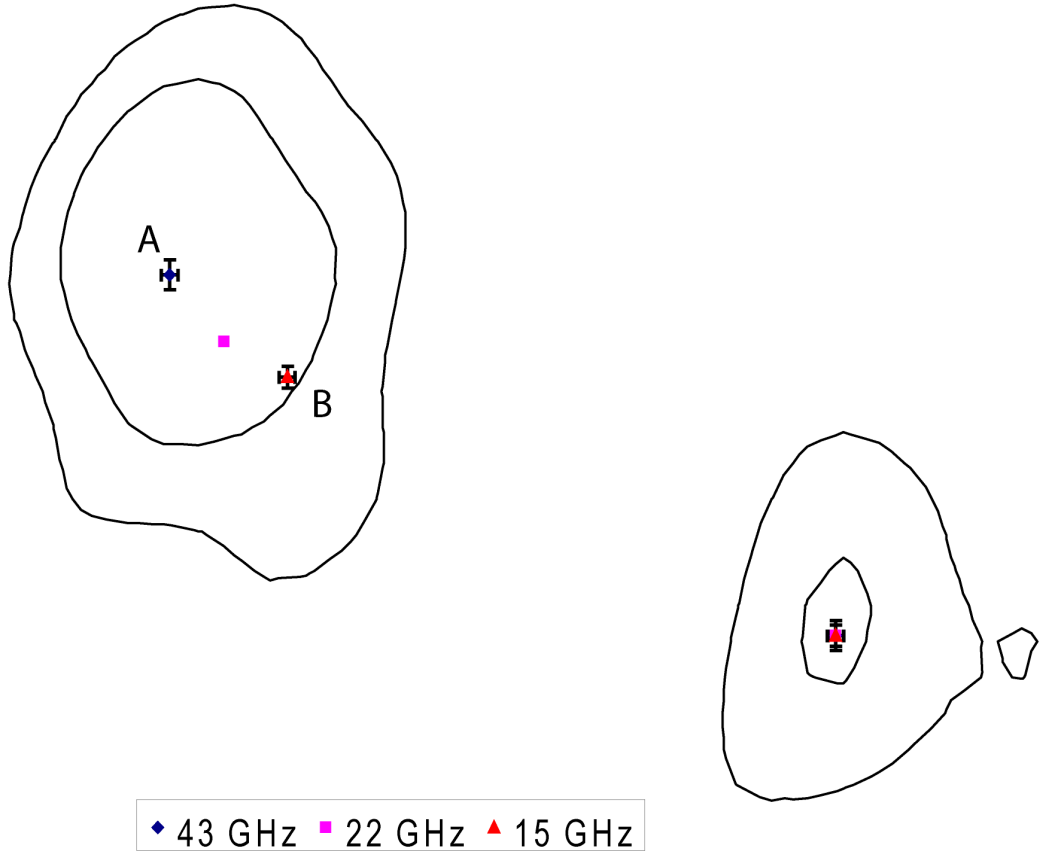}
\caption{An overlay of the frequency dependent component positions
on the 43 GHz image at epoch 2006.07. This figure is similar to
Figure 7, except that the components are re-registered relative to
the 22 GHz secondary position. This tells us how the map center at
one frequency is related to the map centers at the other
frequencies. To get an absolute registration within the 43 GHz radio
image, we place the 3 coincident secondary positions at the location
of the best fit Gaussian to the 43 GHz secondary flux density. The A
and B are labels that show the component locations within the
theoretical two component model. Component A is synchrotron self
absorbed and component B is very steep spectrum.}
\end{figure}
\par The re-registered map in Figure 8 makes the component positions appear almost linear.
We believe that our assignment of errors justifies further analysis
of the frequency dependent component separation data. Apparently, we
have uncovered sub-structure on scales smaller than the 43 GHz
beamwidth. But, unlike the optically thick case in which the core
position movement reflects the location of the frequency dependent
$\tau =1$ surface, the 2006.07 core spectrum is optically thin below
22 GHz and we are revealing direct data on the distribution of the
synchrotron emitting gas. In particular, the spread in the 15 GHz
and 22 GHz core positions in Figure 8, reflects the fact that there
are significant amounts of 22 GHz emitting plasma to the east of the
plasma that emits the 15 GHz data. In fact, the spectrum in Figure 5
and core component positions are well represented by a two component
model. Component A is located $\approx$ 40 $\mu$as to the northeast
of the 43 GHz core position, along a line connecting the three core
position components and component B is $\approx$ 40 $\mu$as to the
southwest of the 15 GHz core position along the same axis. Component
A is synchrotron self absorbed with a peak flux density of $\approx$
27 mJy at $\approx$ 32 GHz. Component B is steep spectrum with
$\alpha \approx 2.4$. We only introduce the model to give a
qualitative feel for the underlying substructure not to fine tune
the parameters. This model explains the inflection point in the SED
at epoch 2006.07 inferred from Figure 5 and predicts $\approx$ 3 mJy
at 43 GHz in component B, similar to what we see in the southwest
core extension in the radio image in Figure 4. Again, we caution
that this feature might just be phase noise. Without a higher signal
to noise detection, we can not say more. However, the spread in
component positions and the SED are consistent with there being some
emission in this region.
\begin{figure}
\includegraphics[width=160 mm, angle= 0]{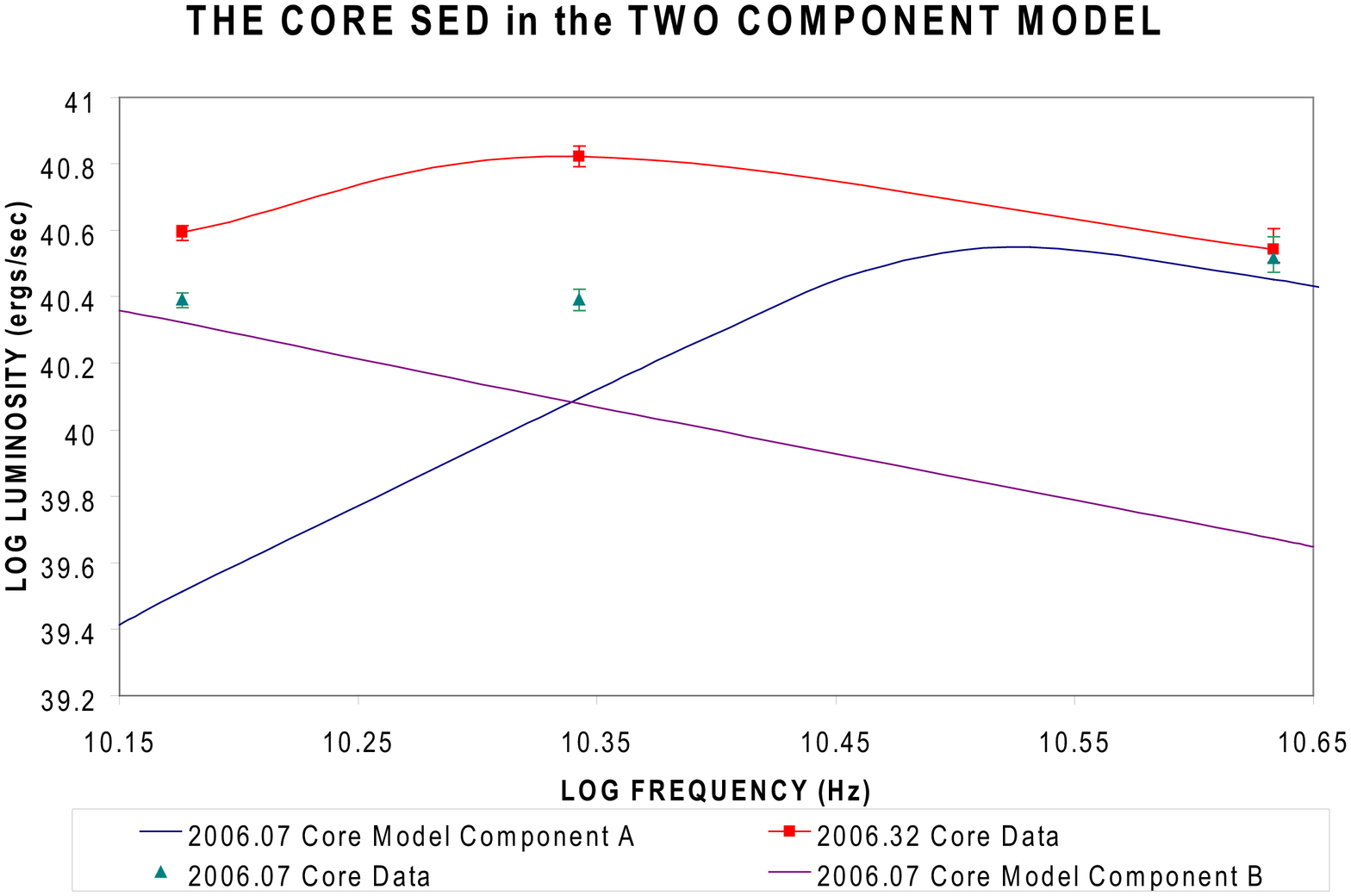}
\caption{Comparison of SEDs of the theoretical two component model
of the core at 2006.07 with the SED at 2006.32}
\end{figure}
\par The reason for pursuing this two component model in Figure 8 is
not to make precise statements about the components that are not
directly detected, but to explore the more general implications of a
core substructure at  2006.07, composed of an optically thick
component and a steep spectrum component, that is required to fit
the SED. To this end, we plot the SED of the theoretical components
A and B relative to the SEDs at 2006.07 and 2006.32 in Figure 9.
\par The first
deduction is that the flare at 2006.32 can not be created by the
simple adiabatic expansion of the flaring core component A at
2006.07. If it were simple adiabatic expansion of a flare that is
peaked at 32 GHz at 2006.07, then the optically thin flux at 43 GHz
should be much weaker at 2006.32, \citep{mof75}, but it is actually
stronger. Thus, the flare at 2006.32 requires a large injection of
energetic particles and energy after 2006.07.
\par The second important implication is that component B is very
steep spectrum above 15 GHz. Figure 9 shows that there is some
mechanism that causes the high energy spectrum to decay rapidly in
the core of Mrk 231. All the components are extremely steep in the
optically thin region, thus there is a very efficient cooling
mechanism that preferentially cools the particles with the highest
energy first. The most common mechanism in AGN that fit this
prescription are inverse Compton cooling from a strong radiation
field or synchrotron emission from a strong magnetic field. The
cooling is so efficient that by the time that an ejection from the
nucleus is far enough away so that it could be resolved by 43 GHz
VLBA, it is generally too weak to be detected. Even with all 10
antennas working at peak efficiency, two good calibrators at 43 GHz
and good weather, partial resolution of the ejecta might be the best
one can do when studying flare evolution in the nucleus of Mrk 231.
\par The next section is the main topic of this study. We explore
the formation of the powerful flare at 2006.32. Furthermore, we try
to get an understanding of the rapid cooling mechanism that causes
drastic temporal changes in the emissivity of the ejected flares.

\section{The Core Flare at Epoch 2006.32}The data in Table 1
indicates that the NE component is highly variable and is likely
associated with the AGN nucleus. The two epochs of the observations, 2006.07 and 2006.32,
reveal a strong flare at 22 GHz that increased the flux density by
45 mJy in just 91 days. The inverted spectrum from 15 GHz to 22 GHz
juxtaposed to the steep spectrum, $\alpha =1.8$, from 22 GHz
to 43 GHz is extremely unusual for a radio source and provides a
strong constraint on the nature of the flare that we exploit in this
section. The steep spectrum above 22 GHz is inconsistent with
free-free emission unless the temperature of the gas is $< 10\,
^{\circ}$ K which is unlikely within a few pc of a quasar nucleus.
Thus, we consider synchrotron radiation as a more likely source of
the emission.
\par There are two possible explanations for the inverted spectrum
near 22 GHz. First, the spectral luminosity, $L_{\nu}$, might be
seen through a free-free absorbing (FFA) screen. For a powerlaw synchrotron
spectrum seen through a FFA gas with number density of free electrons, $N_{e}$, temperature, $T$, and
thickness in parsecs, $L_{pc}$, we adopt the following expressions from \citet{wal00},
\begin{eqnarray}
&& L_{\nu} = L_{0} \times \left(\frac{\nu}{2.2 \times 10^{10}\mathrm{Hz}}\right)^{\alpha}[(1-f)+ fe^{-\kappa(\nu)}]\\
&&\kappa(\nu) = 9.8 \times 10^{-3}L_{\mathrm{pc}} N_{e}^{2}
T^{-1.5}\nu^{-2}[17.7 + \ln (T^{1.5}\nu^{-1})]\;.
\end{eqnarray}
We have only three data points in Table 1, the flux densities at 15
, 22 and 43 GHz of the core. If we rewrite the free-free opacity as
$\kappa(FFA) \equiv \kappa_{o}\nu^{-2}[17.7 + \ln
(T^{1.5}\nu^{-1})]$, we can separate out the frequency dependence,
where the coefficient $\kappa_{o} \equiv 9.8 \times
10^{-3}L_{\mathrm{pc}} N_{e}^{2} T^{-1.5}$. We assume that the
emission region is a homogeneous sphere ejected along the jet
direction toward the secondary. This is certainly more accurate than
a continuous jet for emission that shows no evidence in Figure 5 or
Table 1 of being quasi-steady. The lack of protracted periods of
quasi-steady flux is in contradiction to what would be expected from
a large smoothly varying structure, like the jet models in
\citet{kon81}. Every indication is that the emission around 2006.32
is in the form of a large impulsive burst on the background of some
low luminosity feature (perhaps the continuous jet). The episodic
flaring nature of the core indicated in Section 4, Figure 5 and
Table 1 is consistent with large ejections of highly energized
plasma, with the epoch 2006.32 being the most energetic. The epoch
2006.32 plasmoid is likely oblong and irregular with an
inhomogeneous density, but without any prior knowledge of these
geometric properties, we use a spherical solid as a zeroth order
approximation. With this assumption of a homogenous sphere, the
coefficient $\kappa_{o}$ is a constant in our calculations. We also
assume for now that the covering factor, $f = 1$. For every value of
$T$, we have three variables in equations (5) and (6), $\kappa_{o}$,
$\alpha$ and the normalization $L_{o}$. Thus, we can simultaneously
solve for the three variables as three equations (the 3 measured
flux densities converted into spectral luminosities in equation (5))
with 3 unknowns (the 3 variables) and the resultant will fit the
three flux density data points perfectly. We find the unrealistic
result that $\alpha \approx 3.1$ for the broad range of temperatures
$10^{3}\mathrm{K} < T < 5\times 10^{5}\mathrm{K}$. The extremely
steep background spectrum is a consequence of the fact that the
observed spectrum is very steep just beyond the spectral break,
$\alpha \approx 1.8$ from 22 GHz to 43 GHz. If it is the FFA
absorbing gas that creates the spectral peak between 15 GHz and 22
GHz in Table 1 then it necessarily has a strong effect on the
spectral slope at 22 GHz. In particular, since $\kappa(FFA)$ is a
function of $\nu$, the attenuation is highly significant at 22 GHz.
Consequently, the background synchrotron spectrum at $\approx$ 22
GHz would have to be appreciably steeper than indicated by the
observed value of $\alpha=1.8$.
\par Secondly, the spectral luminosity might be a consequence of
 radiative transfer through a synchrotron self
absorbed (SSA) plasma. The strategy is to perform the final
calculation in the plasma rest frame using known variables from
observation. We designate the observed quantities with a subscript, ``o",
 in the following expressions. Taking the standard result for the
SSA attenuation coefficient in the plasma rest frame and noting that
$\nu = \nu_{o} / \delta$, we find from \citet{rey96,gin69},
\begin{eqnarray}
&& \mu(\nu)=\frac{3^{\alpha +
1}\pi^{0.5}g(p)e^{2}N_{\Gamma}}{8m_{e}c}\left(\frac{eB}{m_{e}c}\right)^{(1.5
+
\alpha)}\nu_{o}^{(-2.5 + \alpha)} \delta^{(2.5 + \alpha)}\;,\\
&& g(n)= \frac{\Gamma[(3n + 22)/12]\Gamma[(3n + 2)/12]\Gamma[(n +
6)/4]}{\Gamma[(n + 8)/4]}\;,
\end{eqnarray}
The Doppler factor, $\delta$, is given in terms of $\Gamma$, the
Lorentz factor of the outflow; $\beta$, the three velocity of the
outflow and the angle of propagation to the line of sight, $\theta$;
$\delta=1/[\Gamma(1-\beta\cos{\theta})]$ \citep{lin85}. This
equation derives from an assumed powerlaw energy distribution for
the relativistic electrons, $ N(\gamma)= N_{\Gamma}\gamma^{-n}$,
where $\gamma$ is the thermal Lorentz factor and the radio spectral
index $\alpha = (n-1)/2$.  The radiative transfer equation was
trivially solved in \citet{gin69} to yield the following parametric
form for $L_{\nu}$ from the SSA source,
\begin{eqnarray}
&& L_{\nu} = \frac{L_{0}\nu^{-\alpha}}{\mu(\nu)} \times \left(1 -
e^{-\mu(\nu) R}\right)\;,
\end{eqnarray}
where R is the radius of the spherical region in the rest frame of
the plasma. We can make the same decomposition that we made for
$\kappa(FFA)$ for the SSA attenuation coefficient,
$\mu(SSA)=\mu_{o}\nu_{o}^{(-2.5 + \alpha)}$. If we assume that the
source is spherical and homogeneous then we have three variables in
(9), $\mu_{o}$, $\alpha$ and $L_{o}$. Since we have three equations
(the 3 measured flux densities converted into spectral luminosities
in equation (9)) with 3 unknowns (the 3 variables), we can fit the
three flux density measurements perfectly. We find that $\alpha =
2.26$ (or $n= 5.52$). Although this is incredibly steep for a radio
source it is still much more reasonable than what we found for FFA.
Thus, we consider SSA a much likelier interpretation of the peak in
the synchrotron spectrum than FFA.
\par Thusly motivated, we proceed to detail the spherical, homogeneous SSA model of the core.
The synchrotron emissivity is given in \citet{tuc75} as
\begin{eqnarray}
&& j_{\nu} = 1.7 \times 10^{-21} (4 \pi N_{\Gamma})a(n)B^{(1
+\alpha)}(4
\times 10^{6}/ \nu)^{\alpha}\;,\\
&& a(n)=\frac{\left(2^{\frac{n-1}{2}}\sqrt{3}\right)
\Gamma\left(\frac{3n-1}{12}\right)\Gamma\left(\frac{3n+19}{12}\right)
\Gamma\left(\frac{n+5}{4}\right)}
       {8\sqrt\pi(n+1)\Gamma\left(\frac{n+7}{4}\right)} \; , \, a(5.52)=0.102\;.
\end{eqnarray}
We can relate this to the observed flux density, $S(\nu_{o})$, in the
optically thin region of the spectrum (43 GHz) using the
relativistic transformation relations from \citet{lin85},
\begin{eqnarray}
 && S(\nu_{o}) = \frac{\delta^{(3 + \alpha)}}{4\pi D_{L}^{2}}\int{j_{\nu}^{'} d V{'}}\;,
\end{eqnarray}
where $D_{L}$ is the luminosity distance and $j_{\nu}^{'}$ is evaluated in the plasma
rest frame at the observed frequency. A second
equation arises from our SSA fit to the data above,
\begin{eqnarray}
&& \mu(\nu = \mathrm{19.5 GHz})R=1\;.
\end{eqnarray}
Since, we already determined $\alpha = 2.26$ then the pair of
equations (12) and (13) with the expansions of equations (7), (8),
(10) and (6.7) indicate that we have two equations in four unknowns,
$R$, $B$, $\delta$ and $N_{\Gamma}$. We explore this solution space
by setting $R$ and $\delta$ fixed then solving (12) and (13) for $B$
and $N_{\Gamma}$, numerically. The results are displayed graphically
in Figure 10.
\begin{figure}
\begin{center}
\includegraphics[width=100 mm, angle= 0]{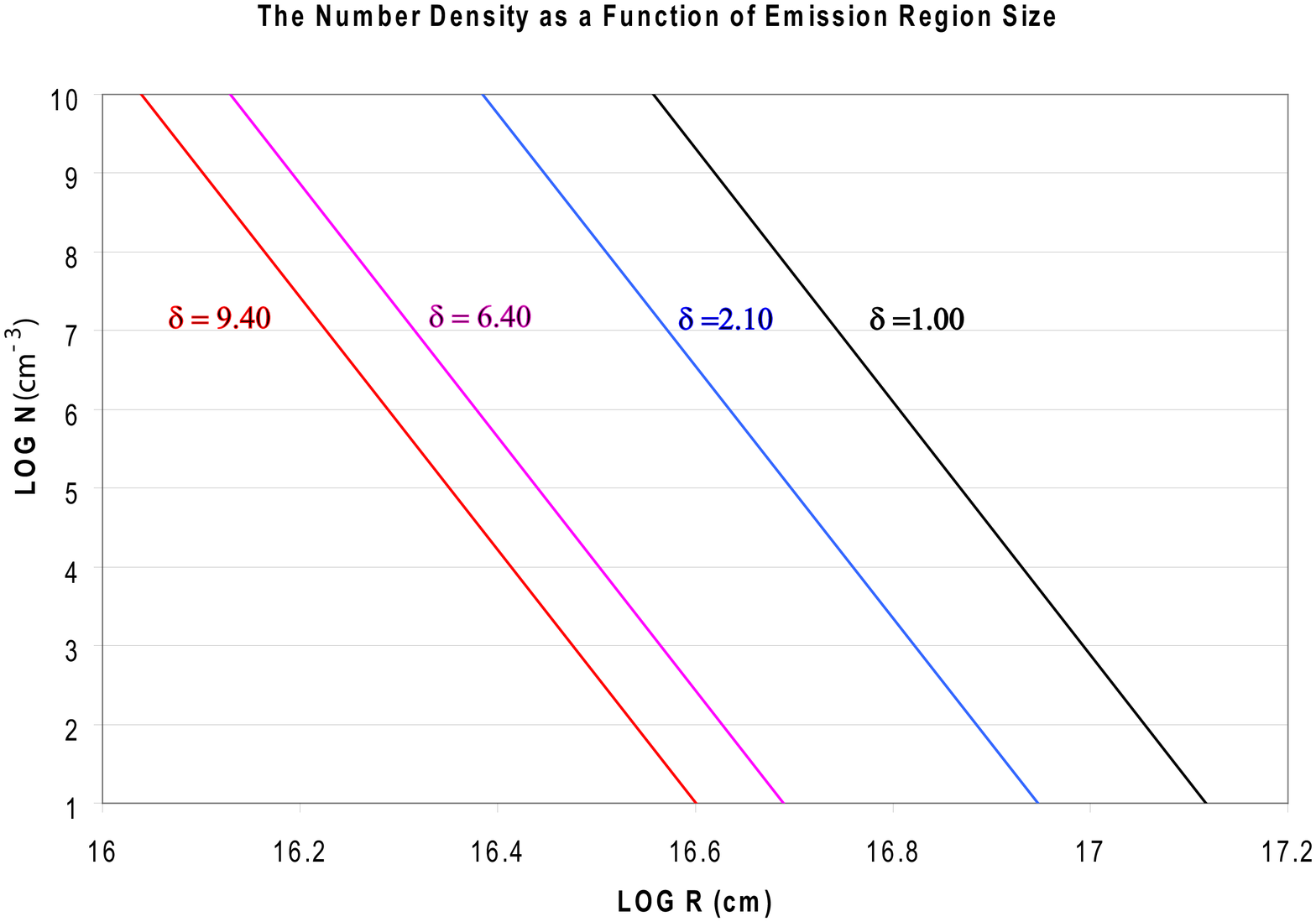}
\includegraphics[width=100 mm, angle= 0]{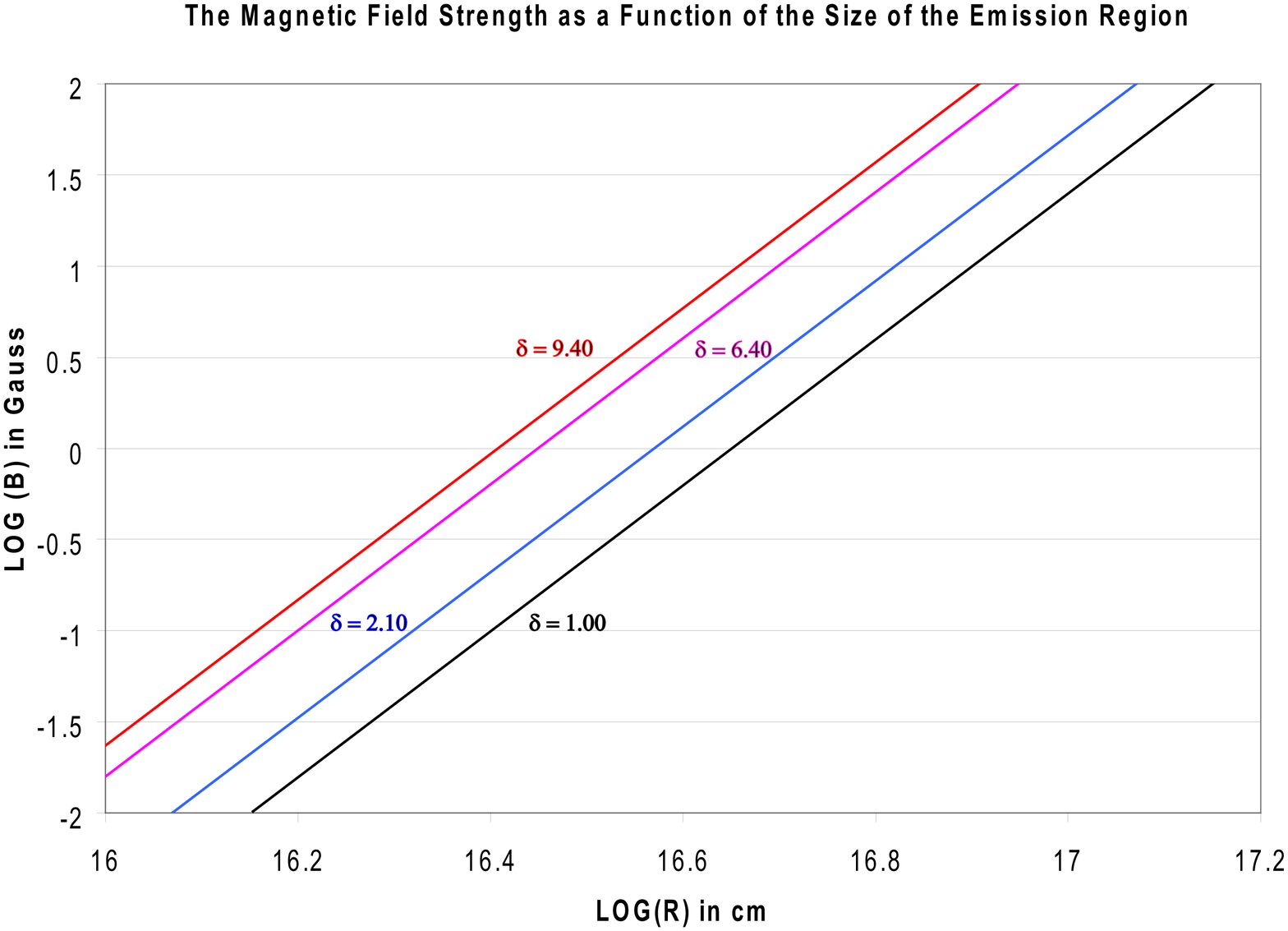}
\caption{The number density
(top) and the magnetic field strength (bottom) in a homogeneous
synchrotron self-absorbed spherical model of the 22 GHz flare at
epoch 2006.32 as a function of Doppler factor, $\delta$ and spherical
radius of the emission region, R.}
\end{center}
\end{figure}
We solve for the actual number density, $N$, from the coefficient
$N_{\Gamma}$ by integrating the powerlaw distribution for
$\Gamma=5.52$ down to $\gamma=1$, $N \approx N_{\Gamma}/4.22 $. We
justify the use of $\gamma = 1$ as a minimum of integration in our
discussion of Table 3 and equation (18) below. We randomly chose
four values of $\delta$ that seem reasonable from VLBI studies of
radio loud quasars, $\delta=1.00$ (non-relativistic), $\delta=2.10$
(mildly relativistic), $\delta=6.40$ and $\delta=9.40$ (highly
relativistic) \citep{kel04,ver94}. For each value of $\delta$ we
found an exact powerlaw solution for both $B$ and $N$. The powerlaw
for $N$ is extremely steep. It is apparent that $N$ and $B$ take on
extraordinary values outside of a narrow range of radii. These radii
are smaller than the 43 GHz beam minor axis - roughly between about
1/3 to 1/30 of the 43 GHz beam minor axis, so the emission region
can not be resolved by VLBA.
\begin{figure}
\includegraphics[width=150 mm, angle= 0]{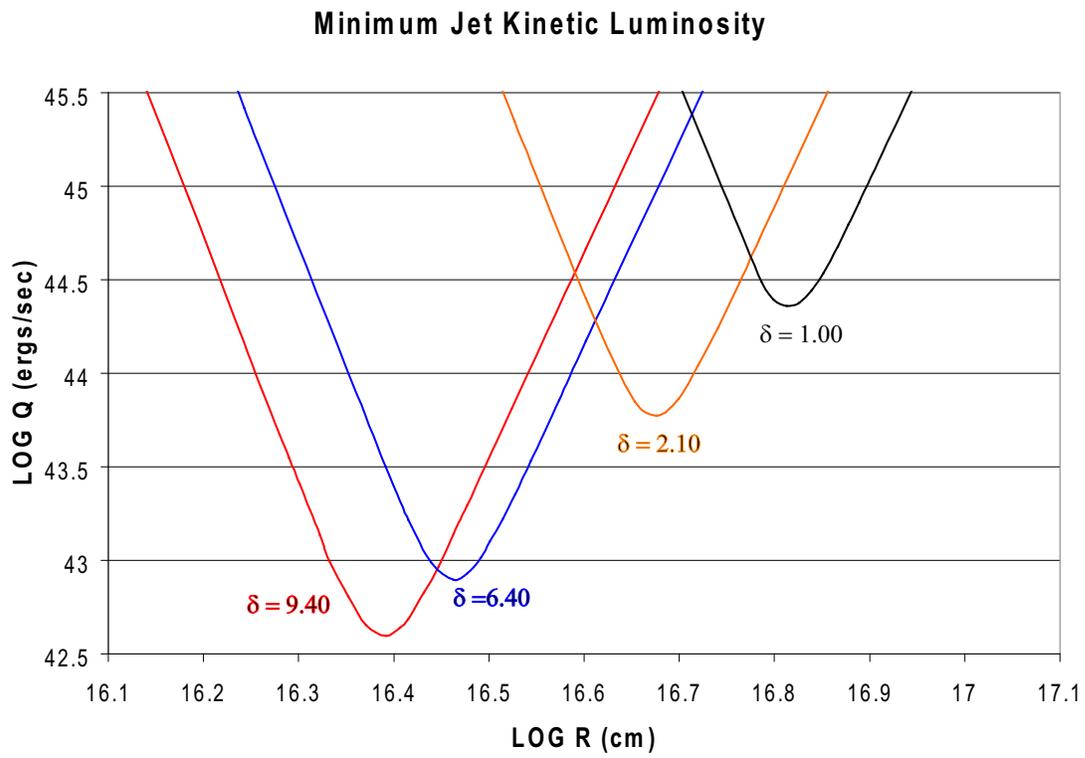}
\caption{The minimum kinetic energy flux, $Q$, for homogeneous
spherical source models as a function of $R$ and Doppler factor,
$\delta$.}
\end{figure}
\par It is straightforward to compute the total energy in a spherical
volume from these solutions for $B$ and $N$,
\begin{eqnarray}
 && E(R) = \int{(U_{B}+ U_{e})}\, dV = \frac{4}{3}\pi R^{3}\left[\frac{B^{2}}{8\pi}
+ \int_{1}(m_{e}c^{2})(N_{\Gamma}\gamma^{-n + 1})\, d\gamma
\right]\;.
\end{eqnarray}
The justification for using 1 as the minimum value of $\gamma$ in
(14) is given in our discussion of Table 3 below. The energy in the
observers frame is $E_{o} = \delta E$. From equation (14), we can
derive a lower bound on the flare kinetic energy flux,
$Q_{min}(R)=\delta E(R)/t_{max}$, where $t_{max}$ = 91 days is the
maximum amount of time that it took the flare to grow (i.e., the
separation in time between our two observations in 2006). We plot
the resulting $Q_{min}(R)$ values in Figure 11 for each of the four
values of $\delta$. Each of the four plots of $Q_{min}(R)$ has a
minimum that we designate as the minimum kinetic energy flux for
that value of delta, $Q_{min}$. To the left (right) of the minimum
in each curve, the flow is thermally (magnetically) dominated. The
minimum values follow an exact functional dependence on $\delta$,
$\log {Q_{min}}= 44.36 -1.8153 \delta$. The properties of these four
$Q_{min}$ solutions are tabulated in Table 3.
\begin{table}
\caption{Details of the Minimum kinetic energy flux Solutions}
{\footnotesize\begin{tabular}{cccccccc} \tableline\rule{0mm}{3mm}
 $\delta$ &  $Q_{min}$ & R & B &  N  & $t_{syn}$ & $T_{b}/T_{max} $ & $\gamma $ \\
 &  ergs/s &  cm & Gauss & $\mathrm{cm}^{-3}$ & $\nu_{o} = 22$ GHz  & & peak \\
 &    &   &    &   &  yrs \tablenotemark{a} & \tablenotemark{b} & \tablenotemark{c}  \\
\tableline \rule{0mm}{3mm}
1.00 & $2.29 \times 10^{44}$ & $6.50 \times 10^{16}$ & 4.43 & $3.35 \times 10^{6}$  & $1.15 \times 10^{-2}$  & 0.86 & $ < 40.7 $  \\
2.10 & $5.94 \times 10^{43}$ & $4.75 \times 10^{16}$ & 2.69 & $ 8.80 \times 10^{5}$  & $1.68 \times 10^{-2}$  & 0.66 & $ < 36.1 $  \\
6.40 & $ 7.86 \times 10^{42}$ & $2.90 \times 10^{16}$ & 1.14& $ 1.81 \times 10^{5}$  & $3.48 \times 10^{-2}$  & 0.47 & $ < 31.7 $  \\
9.40 & $ 3.94 \times 10^{42}$ & $2.47 \times 10^{16}$ & 0.86 & $9.75 \times 10^{4}$  & $4.38 \times 10^{-2}$  & 0.41 & $ < 30.1 $ \\
\end{tabular}}
\tablenotetext{a}{measured in observers frame, see equation (15)}
\tablenotetext{b}{See equations (16) and (17). Based on 68.6 mJy at
15 GHz.} \tablenotetext{c}{see equation (18),
$\nu_{o}(\mathrm{peak}) <$ 22 GHz}
\end{table}
The first column in Table 3 is the value of $\delta$ in the model.
The next column is the minimum kinetic energy flux of the
homogeneous spherical model of the flare. The next 3 columns are R,
B and N for that $Q_{min}$ solution. The synchrotron lifetime
measured in years in the observers frame is calculated in column 6
using the formula from \citet{tuc75},
\begin{eqnarray}
&& t_{syn}\approx (\delta)^{-1} 5 \times
10^{11}B^{-1.5}(\nu_{o}/\delta)^{-0.5} \mathrm{sec}\;.
\end{eqnarray}
We compute this at 22 GHz in the observer's frame in order to see if the very
steep spectrum from 22 GHz to 43 GHz in Table 1 can be explained by
synchrotron cooling - spectral ageing. All of these lifetimes are
short compared to 0.25 years between observations and since the
flare probably expanded from smaller dimensions, $B$ was probably
significantly larger in early stages of the flare, so the
synchrotron cooling rate was probably larger than indicated by
column 6 in the early stages of the flare evolution. Thus, we
conclude that synchrotron cooling is a viable explanation of the
very steep spectrum, high frequency tail of the core spectrum in
Figure 5.
\par The next column tabulates the brightness
temperature, $T_{b}$, if the core flux came from the spherical
volume of radius, R, using the definition from \citet{sch68}
\begin{eqnarray}
&& T_{b}= \frac{S_{\nu}}{\Omega}\frac{\mathrm{c}^{2}}{2
\mathrm{k}_{b} \nu^{2}}\;.
\end{eqnarray}
The values approach the inverse Compton limit that is given in
\citet{mar79} as (see also equations (19) and (20), below)
\begin{eqnarray}
&& T_{max} \sim 10^{12}\delta^{1.2}\,\mathrm{K}\;.
\end{eqnarray}
The large $T_{b}$ values in Table 3 indicate that significant
cooling of the relativistic electrons from inverse self-Compton
losses are working in consort with synchrotron cooling to create the
steep spectrum \citep{kel69,rea94}.
\par The last column in Table 3 is an estimate of
electron energy $\gamma$ at the peak frequency. We know that the
peak is less than 22 GHz, so these values are an upper bound. Using
the formula form \citet{tuc75}, we compute the electron energy at 22
GHz to form an upper bound,
\begin{eqnarray}
&& \gamma_{\mathrm{peak}} \approx
\frac{\nu_{o}(\mathrm{peak})/\delta}{(3 \times 10^{6}B)^{0.5}} \;.
\end{eqnarray}
These are very low values for a relativistic jet and suggest that
the energy spectrum likely goes down to $\gamma = 1$ (i.e., there
are electrons that radiate as low as 20 MHz, but the emission is not
seen due to SSA). The low $\gamma$ values are related to the fact
that all these minimum kinetic energy flux solutions are very highly
magnetized.
\par The most revealing entry in Table 3 is the high value of $T_{b}$. The ratio of inverse Compton
luminosity, $L_{ic}$, to synchrotron Luminosity, $L_{synch}$ was calculated in \citet{rea94} as
\begin{eqnarray}
&& \frac{L_{ic}}{L_{synch}} = \left(\frac{T_{b}G(\alpha,
z)^{1/5}}{10^{12.22}}\right)^5 \left[1 + \left(\frac{T_{b}G(\alpha\,
, \; z)^{1/5}}{10^{12.22}}\right)^5 \right]\;,\\
&& G(\alpha\, , \; z) =\\ \nonumber
&&\left[\frac{v_{\mathrm{op}}/\delta}{3.5\, \mathrm{GHz}}+
\frac{v_{\mathrm{op}}/\delta}{(1+z)1\, \mathrm{GHz}}\times
\left[\left(\frac{v_{\mathrm{high}}}{(1+z)v_{\mathrm{op}}}\right)^{(1 +\alpha)}-1
\right] \right] f_{3}(\alpha)^{-4}(1 + z)^{4}\;,
\end{eqnarray}
where $v_{\mathrm{op}}= 19.5$ GHz is the peak in the observed SSA
spectrum, $v_{\mathrm{high}}=43$ GHz is the highest frequency in the
power law in the observers frame (typically taken as 100 GHz in the
rest frame of the plasma). Our choice of 43 GHz is conservative and
$L_{ic}/L_{synch}$ might be slightly underestimated by this
assumption. This expression also requires the coefficient,
$f_{3}(\alpha)$, that is given by equation (15c) of \citet{sch68}.
For the minimum kinetic energy flux solution with $\delta = 1$ in
Table 3, the data in Table 1 and equations (19) - (20) imply that $
L_{ic}/L_{synch} \approx 2 \times 10^{8}$. The X-ray observations
from 0.2 keV - 50 keV from \citet{bra04} that indicate a total
intrinsic X-ray luminosity of $L_{X}\sim 10^{44}\mathrm{ergs/s}$
from the sum of accretion disk corona luminosity and the jet
luminosity. Combining this with our measured radio luminosity of the
core in Table 1, we observe that
\begin{eqnarray}
&& \frac{L_{ic}}{L_{synch}} \sim \frac{10^{44}
\mathrm{ergs/s}}{10^{41} \mathrm{ergs/s}}= 10^{3}\;.
\end{eqnarray}
 Thus, the $\delta=1$
minimum kinetic energy flux solution is inconsistent with the value
of $L_{ic}/L_{synch}$ determined by observation.
\par One can investigate other possibilities in the $\delta=1$ solution space to see
if a solution that is compatible with equation (21) is viable.
Essentially, $R$ must be increased in order to lower $T_{b}$.
Increasing the spherical radius of the $\delta=1$ solution so that
$T_{b}$ satisfies equation (21) requires that $R > 1.2 \times
10^{12}$ cm. According to Figure 11, this would correspond to $Q
\sim 10^{46}$ ergs/s, the power of the strongest known FR II radio
jets \citep{wil99}. These seems unlikely for a radio quiet quasar.
\par Alternatively, if we
look at the $\delta=9.40$ solution at $R = 3.12\times 10^{16}$ cm we
get an intrinsic value of $T_{b}^{'}= T_{b}/\delta= 3.7 \times
10^{11}\, \mathrm{K}$ \citep{rea94}. Inserting this into equations
(19) and (20), implies that $L_{ic}/L_{synch} \lesssim 10^{3}$ for
this solution in agreement with the bound set by observation in
equation (21). From Figure 11, this solution is magnetically
dominated and $Q \sim 3.5 \times 10^{43}$ ergs/s. This value of $Q$
seems like a more reasonable value for a burst in a radio quiet
quasars, on the order of the long term time averaged $Q$ in a strong
FR I radio galaxy \citep{wil99}.

\par In summary, we argue that the $\delta =1.00$ solution is not physically reasonable on 2
counts:
\begin{itemize}
\item $T_{b}$ is right below the Compton limit for the minimum kinetic energy flux solution.
If this were true then there should be a strong intrinsic X-ray
source (and it should diminish the radio emissivity as well)
\citep{kel69}. Extensive deep observations and very hard X-ray
observations indicate that the intrinsic X-ray luminosity,
$L_{X}\approx 1 \times 10^{44}\mathrm{ergs/s}$ \citep{bra04}. This
is actually at the low end for the intrinsic X-ray luminosity
distribution from the corona of the accretion disk of a quasar which
has the thermal luminosity of Mrk 231 that was noted in the
introduction, $L_{bol}\gtrsim \times 10^{46}\mathrm{ergs/s}$
\citep{lao97}. Thus, the strong intrinsic radio emission and the
relatively weak intrinsic X-ray emission (by radio quiet quasar
standards) is not consistent with $T_{b}$ near the Compton limit.
Alternatively, $R$ can be enlarged, but this presents different
concerns as noted in the second point.
\item The minimum flare kinetic energy flux is $Q_{min}= 2.30 \times 10^{44}\mathrm{ergs/s}$. This
is an Fanaroff Riley II (FR II) kinetic energy flux. The maximum FR
I kinetic energy flux cutoffs at about $Q= 5 \times
10^{43}\mathrm{ergs/s}$ \citep{wil99}. Also notice that the kinetic
luminosity rises very quickly away from the minimum in Figure 11, so
this is a very loose lower bound on $Q$. In order to alleviate the
first concern on $T_{B}$, above, a significant departure from the
minimum kinetic energy flux solution is in order as discussed in
regards to equation (21) and therefore a marked increase in the
flare kinetic energy flux. It seems unlikely (but not impossible)
that a radio quiet quasar should have flares with a strong FR II
kinetic energy flux.
\end{itemize}
\par Based on these two considerations, it seems difficult to reconcile
the large kinetic energy flux with a Doppler factor of order unity
in Mrk 231. Table 3 and the discussion of equations (19) - (21)
indicates that a significant Doppler factor would ameliorate these
concerns. It would also explain the gap between the core and the
secondary in Figures 3 and 4. Small deviations in the line of sight
can greatly alter the flux density since by equation (12), $S_{\nu}
\sim \delta^{5.26}$. Thus, small changes in the velocity or line of
sight (jet wobble) can cause the jet emission between the core and
secondary to be of such low surface brightness that it is too faint
to be detected with VLBI. Similarly, a strong dependence on $\delta$
is compatible with the extreme flares and spectral variations that
appear in Table 1. It seems likely that the subparsec scale radio
jet in Mrk 231 is highly Doppler enhanced, i.e. relativistic and
viewed pole-on. This is not unexpected since numerous BALQSOs have
been determined to be viewed in a pole-on orientation based on radio
variability $T_{b}$ values above the Compton limit
\citep{zho06,gho07}. In particular, LoBALQSOs (those BALQSOs that
exhibit low ionization resonant absorption troughs) seem
inordinately common in these polar samples (1/3 - 1/2 of the polar
BALQSOs). Mrk 231 is a LoBALQSO.
\par There is strong corroboration in the literature for our deduction that Mrk 231
hosts a polar relativistic jet. A $134 \pm 38$ mJy flux density
variation in 1 day at 22.2 GHz was reported in \citet{mcc78}. Using
the methods of \citet{gho07}, the time variablity brightness
temperature is $T_{b} = (12.4 \pm 3.5)\times 10^{12}$ K. In order to
avoid the inverse Compton catastrophe, the analysis in \citet{gho07}
indicates that the line of sight to the jet is less than
$\theta_{max}= (25.6^{\circ})^{+3.2^{\circ}}_{-2.2^{\circ}}$.
\par We also note the fortuitous circumstance that 8.33 GHz VLBA
observations were performed 0.35 years after the strong flare. epoch
2006.67 (BA80B). The historic high levels of 8 GHz of core flux
density at 2006.67 indicated in table 1 and figure 5 is supportive
of the flare evolving to lower frequency 0.35 years later.
Unfortunately we could not recover the 15
GHz data at 2006.67 which would have told us much more.

\section{The Nature of the Secondary Component}In this section we study the interaction of the
secondary component with the surrounding medium. It was shown in
\citet{ulv99}, using low frequency (1.4 GHz - 15.3 GHz) VLBA
observations that the compact radio double is seen through a FFA
shroud. The FFA absorption creates a spectral turnover near 6 GHz.
From Table 1, it is clear that the radio emission of the compact
double is dominated by the secondary at low frequency. Thus, the
emission measure (see equation (6), $EM = \int N_{e}^{2}\, dl $),
that is calculated in \citet{ulv99} from this 6 GHz peak in the
spectrum represents the shroud of gas that envelopes the secondary
and might not be indicative of the gas density surrounding the core.
In \citet{ulv99}, it was stated that $EM \approx 1.2 \times
10^{8}\mathrm{cm}^{-6}\mathrm{pc}$ produced the spectral turnover.
However, it was not clear what temperature was assumed for the gas.
Thus, we re-derive this result in a form that incorporates the
temperature.
\par In order to study the spectral turnover, we want to use the lowest frequencies possible in order
define the shape of the spectral peak accurately. The compact parsec
scale double discussed thus far in this paper appears as an
unresolved ``core" at low frequency. The ``core" is part of a triple radio
source at low frequency. Components to the north and south of the
core are detected at 8.4 GHz and below. At 2.3 GHz, the resolution
is not sufficient to avoid confusion between the ``core" flux and
flux from the north/south extended components. Thus, as a
compromise, we chose data at a resolution matched to the 5 GHz
observation. We fit the data using equation (5) and a background
spectral index of $\alpha =1.7$ for the synchrotron source. Our best
fit used a covering factor, $f=0.98$. The corresponding solution for
the emission measure is close to the value stated in \citet{ulv99},
\begin{eqnarray}
&& EM \approx 9 \times 10^{7}\mathrm{cm}^{-6}\mathrm{pc}
\left(\frac{T}{10^{4}\,\mathrm{K}}\right)^{1.5} \; , \, 2\times
10^{3}\,\mathrm{K}< T < 5\times 10^{4}\,\mathrm{K}\;.
\end{eqnarray}
\par Considering that the secondary is surrounded by a dense cloud of free electrons as indicated in
equation (22) and the secondary does not move appreciably according
to Table 2, even though the core jet is relativistic, it is natural
to consider the possibility that the secondary is being braked to a
slow speed by the surrounding gas that supports $N_{e}$. In physical
language, is the pressure of the secondary in dynamic equilibrium
with the ram pressure of the surrounding gas? To explore this
possibility we first try to estimate the pressure within the
secondary.
\par To estimate the pressure, we first find the $Q$ associated with the secondary and use the
physical size to convert it to a pressure. Since the secondary is
very steep spectrum at all epochs and is mildly variable, we
consider it to be a compact analog of the radio lobes that are seen
in classical double radio sources \footnote{One might also think of
this feature as a strong knot in a jet that disappears, then
reappears as a diffuse low frequency large scale elongated structure
(that is depicted by the low frequency VLBA observations in
\citet{ulv99}) that is barely detected with 8.4 GHz VLBA at a PA
$65^{\circ}$ offset from the sub-pc scale jet direction. The very
steep spectral index in figure 6, above 8.4 GHz (which equates to a
background spectral index of $\alpha = 1.7$ in the FFA fit above),
seems more indicative of lobe emission than knot emission. For such
a steep spectrum structure, the distinction of knot or lobe is
probably irrelevant with respect to the application of estimate of
the jet kinetic energy flux in equation (23)}. Furthermore, this
compact lobe is seen through the effects of a FFA shroud and we can
not detect its low frequency emission, even though it is plausibly
$> 1$ Jy considering the 8.4 GHz flux density in Table 1. Perhaps
the most sophisticated estimate of jet kinetic energy flux was
developed by \citet{wil99} which incorporates deviations from the
overly simplified minimum energy estimates. The method also
considers geometric effects, filling factors, protonic contributions
and low frequency cutoff. The long term time average of the jet
kinetic energy flux of a steep spectrum radio lobe, $\overline{Q}$,
is estimated in \citet{wil99}:
\begin{eqnarray}
&& \overline{Q} \approx 3.8\times
10^{45}L_{151}^{6/7}\mathrm{ergs/s}\;,
\end{eqnarray}
where $L_{151}$ is the total optically thin luminosity from the
lobes in units of $10^{35} \mathrm{ergs/Hz/sr}$ at 151 MHz. The
obstacle to implementing equation (23) is that it depends on the 151
MHz flux density. This is problematic since the 151 MHz flux from
the secondary is blocked from our view by the FFA shroud. The lowest
frequency measurement that we have in Table 1 is 8.4 GHz and we need
to extrapolate this value to 151 MHz. We know that the spectrum is
very steep between 8.4 GHz and 15 GHz, $\alpha = 1.38$, even with
some diminished flux at 8.4 GHz from FFA. There are very few (if
any) known radio lobes that are this steep between 151 MHz and 8.4
GHz. Thus, there is no justification for extrapolating the 8.4 GHz
flux density with $\alpha = 1.38$. A more typical value for steep
lobe emission seems to be $\alpha \approx 0.8$ \citep{kel70,her92}.
Under this assumption, we used Table 1 and equation (23) to estimate
the long term time average kinetic energy flux of the putative jet
connecting the core to the secondary, $\overline{Q} \approx 9.9
\times 10^{41} \mathrm{erg/s}$.
\par Even though this estimate of $\overline{Q}$ is crude, we can use it to estimate the pressure in
the secondary. We can approximate the energy flux through the jet as
$\overline{Q}\approx (1/4) \pi D^{2} v_{adv} U$, where D is the diameter of
the lobe, $v_{adv}$ is the velocity that the lobe is advancing into
the external medium and $U = U_{e} + U_{B}$ is the total internal
energy density. Since in a relativistic gas the pressure $P_{lobe} =
(1/3)U$, we estimate
\begin{eqnarray}
&& P_{lobe} \approx \frac{4\overline{Q}}{3 \pi D^{2} v_{adv}}\;.
\end{eqnarray}
If the lobe is confined by the ram pressure of an external medium moving
with a velocity, $v_{ext}$, and density, $m_{p}N_{ext}$, then we
compute the pressure balance at the leading edge of the lobe to find
\begin{eqnarray}
&& P_{lobe} \approx \frac{4\overline{Q}}{3 \pi D^{2} v_{adv}}\approx m_{p}N_{ext}(v_{adv} - v_{ext})^{2} \;.
\end{eqnarray}
If we implement a fiducial distance of 1pc and  assume that the
medium is fully ionized then by equation (22), $N_{ext} \approx
10^{4} \mathrm{cm}^{-3}$. We also get a value of $D= 7.70 \times
10^{17}\mathrm{cm}$ from the angular size of the emission region,
0.31 mas (this was determined from model fits to the 15 GHz VLBA
data by \citet{ulv00} and was verified by our fits in various epochs
as well). We can then solve (25) with these parameters derived from
the FFA fit and the VLBA data to get a qualitative understanding of
the ram pressure confinement of the lobe. We explore equation (25)
in two limits, first the enveloping gas is stagnant $v_{ext}= 0$
which yields $v_{adv} = 0.012c$ = 3490 km/sec. Secondly, we note
that the distance from the secondary to the core is typical for the
location of the LoBAL gas. The velocity of the MgII absorption
trough relative to the H$\alpha$ emission (which we take as an
approximation to the quasar rest frame) from \citet{smi95} can be
equated with the velocity of the enveloping medium, $v_{ext}\approx
0.0173c$, from which equation (25) implies that $v_{adv} = 0.025c$ =
7570 km/sec. These values are consistent with the slow advance speed
indicated in Table 2. The dynamic estimates correspond to a
separation increase between the core and the secondary in 9.4 years
of $0.041\sin{\theta}$ mas and $0.089\sin{\theta}$ mas, respectively
(where $\theta$ is the angle between the line of sight and the jet
propagation axis). These values are consistent with our observations
which by equation (4) bounds $v_{adv} < (0.053/\sin{\theta})c$, at
the one sigma level.
\par In summary, the relativistic jet is over-pressurized relative to
the gas density of the FFA absorbing gas, so it advances. However,
the FFA gas is very dense and the kinetic energy flux is modest,
thus the jet can drill through the dense medium relatively slowly at
a non-relativistic rate. Independent of the assumptions involved in
deriving and evaluating equation (25), this qualitative conclusion
is robust and is consistent with the observed VLBA component
separation data in Table 2.
\par This analysis raises the question, what is the medium that is
confining the relativistic jet and braking its velocity to a virtual
halt? Is it the putative dusty molecular torus, $v_{ext}= 0$, or is it
the BAL gas, $v_{ext}= v_{BAL}$? We argue that the FFA absorbing
screen is more likely the BAL gas than the molecular torus based on
the following points.
\begin{itemize}
\item The low frequency VLBI observations of \citet{lon03} led them
to conclude that ``the lack of 21 cm absorption or maser
amplification against the compact core (which is the compact double
with our higher resolution) shows that the molecular disk does not
cover the central source."
\item The high Doppler factor inferred from our modeling of the core
flare indicates that the jet is viewed in a nearly pole-on
orientation. Furthermore, the existence of BAL troughs in the
spectrum of Mrk 231 indicates that our line of sight is through the
BAL gas. Thus, the jet direction seems pass through the BAL gas on
pc scales.
\end{itemize}
In summary, the observational data is consistent with the jet
ramming through the BAL gas and inconsistent with the jet plowing
through the molecular disk. Thus, we proceed to consider the BAL gas
as the FFA medium.
\par If the jet is ramming through the BAL gas then the jet can be
used as a remote probe of properties of the BAL region. Thus, MRK
231 might be uniquely suited for studying the BAL gas. We find that
the properties of the gas derived in this section are consistent
with published interpretations of BALQSO physics. The main piece of
information is the emission measure in equation (22), that does not
have a unique decomposition in terms of the number density
distribution $N_{e}(r)$ and the extent of the FFA region, $L$. In
\citet{ulv99}, they chose $L = 1 \mathrm{pc}$ and a constant $N_{e}
= 1.1 \times 10^{4} \mathrm{cm}^{-3}$. We interpret this value of
$L$ within the context of the polar models of BAL outflows flows in
\citet{pun99,pun00,pro04}. Consider a conical outflow that
experiences very little acceleration as might be the case in the
LoBAL region (near UV lines) since most of the line driving
acceleration is provided by far UV and EUV absorption lines
\citep{mur95}. In this scenario, by conservation of mass, $N_{e}(r)
= N_{o}/r^{2}$ and we can integrate the emission measure in general
form as
\begin{eqnarray}
&& EM =\int_{L}^{\infty} N_{o}^{2}r^{-4}dr
=\frac{1}{3}(N_{e}(r=L))^{2}L\;, L = L_{obs}/\sin{\theta}\;,
\end{eqnarray}
where the observed $L_{obs}$ is the projection of $L$ in the sky
plane and $\theta$ is the angle to the line of sight to the radio
jet axis (symmetry axis of the BAL outflow). In \citet{pun00}, it
was calculated that $\theta =15 \,^{\circ}$ was typical for a LoBAL
or from Table 2, $L = 3.75$ pc and from equations (22) and (25),
$N_{e}(r=L) \approx 4.9 \times
10^{3}\mathrm{cm}^{-3}\left(\frac{T}{10^{4}\,\mathrm{K}}\right)^{0.75}
$. If we assume that the absorbing column is almost completely
ionized ($N_{e}\approx N$) in the BAL region,
 as deduced in \citet{ham98}, then we can estimate the total hydrogen column,
\begin{eqnarray}
&& N_{H}= \int^{\infty}_{L}N(r)dr \approx 5.6 \times
10^{22}\left(\frac{T}{10^{4}\,^{\circ}\,\mathrm{K}}\right)^{0.75}
\mathrm{cm}^{-2}\;.
\end{eqnarray}
The total hydrogen column is sufficient to make the BAL absorption
lines optically thick to resonant absorption \citep{mur95,ham98}.
Superficially, one would think that the value of $N_{H}$ in equation
(27) is incompatible with the BAL gas in Mrk 231. This is actually
not the case as discussed at length in \citet{ham98,ham03}. It was
found that many BALQSOs show absorption in PV $\lambda \lambda$1118,
1128 with an optical depth $\geq 0.2$. For such a rare species this
is impossible, unless the CIV $\lambda \lambda$1548, 1550 troughs
are severely saturated. After studying many high resolution spectra
of BALQSOs, \citet{ham03} concluded that over half of the BALQSOs
with deep far UV spectra showed significant PV $\lambda
\lambda$1118, 1128 absorption. The implication was that these
BALQSOs must have CIV $\lambda \lambda$1548, 1550 troughs that are
completely saturated with the bottoms of the troughs filled in by
scattered light and un-attenuated (or mildly attenuated) lines of
sight that are a consequence of partial covering of the central AGN
by the predominant thick absorber. After, numerically modeling the
absorption region, \citet{ham98} found that the CIV $\lambda
\lambda$1548 1550 troughs are saturated as a consequence of
absorption presented by a column that contained almost completely
ionized hydrogen, $10^{22}\mathrm{cm}^{-2}< N_{H} <
10^{24}\mathrm{cm}^{-2}$. The upper limit is set by the fact that
the gas must be optically thin to Compton scattering. It was
concluded that these large column densities are likely common in BAL
regions. Based on the analysis of \citet{ham98,ham03}, the values of
$N_{H}$ determined by equation (27) are typical of many BALQSO
absorbing columns. Furthermore, we remind the reader that there is
very little neutral hydrogen absorption towards the nuclear
secondary as noted in \citet{lon03}. Yet by equations (22) and (27)
there are large values of EM and $N_{H}$ which implies that the
intervening hydrogen column density must be almost completely
ionized as expected from the ionization models of dense BAL winds in
\citet{ham98}. In particular, the photo-ionization models of
\citet{ham98} yield a range of allowed ionization states: for a
total hydrogen column density of
$N_{H}=6\times10^{22}\mathrm{cm}^{-2}$, the neutral hydrogen column
density is quite small, $5 \times 10^{16}\mathrm{cm}^{-2}< N_{HI} <
5 \times 10^{17}\mathrm{cm}^{-2}$.

\par From our analysis above, we have enough parameters to estimate the
kinetic energy flux of the the BAL wind, $Q(BAL)$ in terms of the
parameters of the enveloping gas surrounding the leading edge of the
secondary. First, assume that the BAL wind has a bulk velocity in
the range of velocity spread given by the MgII BAL absorbing gas and
NaI D absorbing gas relative to the quasar rest frame, $0.0151c <
v_{BAL} <0.0173c$. From the numerical models in \citet{pun00}, a
polar LoBAL wind will have an opening angle of $\approx 20^{\circ}\,
- \,30^{\circ}$. Combining this opening angle and the BAL wind
velocity with the 3.75 pc displacement from the nucleus and
$N_{e}\approx 4.9 \times 10^{3}\mathrm{cm}^{-3}$ yields $2 \times
10^{43}\mathrm{ergs/s}< Q(BAL) < 6 \times 10^{43}\mathrm{ergs/s}$.
It is curious that this is similar to the kinetic energy flux of the
relativistic flare that was discussed in section 5. One can compare
this parametric analysis, with the theoretical polar BAL models in
Table 1 of \citet{pun00}. First of all, for the numerical models,
the kinetic energy flux in the BAL wind solutions is $7.2 \times
10^{42}\mathrm{ergs/s}< Q_{\mathrm{theoretical}}(BAL)< 2.3\times
10^{44}\mathrm{ergs/s}$, in agreement with what we deduced for MRK
231. However, the mass density in the BAL wind of Mrk 231 is
estimated to be a factor of 10 - 50 times larger than that which is
predicted by mass conservation in the numerical models. A possible
explanation of the additional gas density is that the polar BAL
entrains gas as it propagates through the surrounding medium. This
seems reasonable on two counts. First of all, there is the dusty,
enveloping gas that partially obscures the nuclear region and
reddens the spectrum. This gas distribution could be a consequence
of the merger that is in progress with a nearby smaller galaxy to
the south \citep{arm94}. So there is a potential for significant gas
entrainment. Secondly, the BAL absorption is deep, but not
particularly blue-shifted for a BALQSO, only about 5000 km/s: this
is consistent with a resistive drag resulting from mass loading of
the wind.

\section{Discussion}
In this article, we used high frequency VLBA observations to
determine that the central engine of the nearby BALQSO Mrk 231,
ejects relativistic plasma along a trajectory close to the line of
sight. Furthermore, it appears that this jet is interacting strongly
with the dense BAL gas at a de-projected distance $\sim$ 3 - 4 pc
from the radio core.
\par The most striking finding was the strong
22 GHz flare that emerged from the core between epochs 2006.07 and
2006.32. The core spectrum, during the flare, is characterized by
extreme gradients, namely it transitions from being very steep
spectrum, $\alpha \approx 2$ above 22 GHz to being inverted $\alpha
< 0$ at $\approx 15$ GHz. The magnitude of the flare combined with
the three month interval between observations sets a lower bound on
the time variability brightness temperature of $T_{b}
> 2\times 10^{10}\,\mathrm{K}$, below the inverse Compton
limit, $\sim 10^{12}\,\mathrm{K}$. However, all attempts in section
5 to model the high frequency peak of the spectral turnover, 19.5
GHz, in combination with the steep spectral index above 22 GHz
($\alpha \approx 2$) indicate that the flare is synchrotron
self-absorbed and $T_{b} \approx 10^{12}\,\mathrm{K}$, unless the
flux density is Doppler boosted. Our Doppler boosted models indicate
a kinetic energy flux, $Q \sim 3 \times 10^{43} \mathrm{ergs/sec}$
and an intrinsic (rest frame of the plasma) brightness temperature
$\gtrsim 10^{11}\,\mathrm{K}$.
\par Our discovery of a relativistic jet in a radio quiet quasar does not stand alone. There are the
aforementioned BALQSOs in \citet{gho07,zho06} as well as PG 1407+263
in \citet{blu03} and III Zw 2 in \citet{bru00}. By definition, most
radio quiet quasars are weak radio sources, so there are not many
observations with VLBI. Consequently, we can not say if the
relativistic jets that have been detected are more the exception or
the rule.
\par The compact nature of the jet emission in MRK231 and the strong interaction with
the surrounding medium has many strong similarities to GHz peaked
(GPS) radio sources \citep{ode98}. In the interval between 2006.07
and 2006.32, Mrk 231 seemed to transform from a weak version of a
GPS source (as it was in 1996.93) to a weak version of a rare class
of radio sources known as high frequency peakers, but on a scale
1000 times smaller \citep{ori08}. It is weakly polarized like GPS
sources, but unlike GPS sources Mrk 231 is highly variable
\citep{ode98}. We were unable to reliably detect any morphological
changes associated with this change in character. It would be
interesting to see if long term VLBA monitoring at 43 GHz and 22 GHz
would reveal ejecta emanating from the core $\sim$ a month after a
flare. Another interesting aspect of high frequency VLBA monitoring
would be to understand the flare growth and decay time scales. The
information gleaned from this data would further constrain the
physical model of section 5. This monitoring would undoubtedly
reveal large time variability brightness temperatures and might
confirm the presence of synchrotron cooling, i.e. is there a short
lived 43 GHz precursor to a 22 GHz flare. If the basic SSA model of
the flare that was presented in section 5 is confirmed by these
observations, a natural question arises: what is the origin of the
strong magnetic field , $\sim$ a few G, on scales approaching
$10^{17}$ cm? If the field propagates from the base of the jet, does
it tell us something about the turbulent plasma inside the accretion
disk?
\par Another important finding was that the core was unpolarized even during a flare. This implies that
the magnetic field is not ordered or there is a huge amount of
weakly magnetized gas around the central engine which depolarizes
the synchrotron emission by Faraday rotation. The arguments in
section 5, regarding $T_{b}$ seemed to favor magnetically dominated
models. It is hard to understand how a strong field would not be
ordered in a magnetically dominated plasma. This would seem to
indicate a dense magnetized envelope surrounding the radio core.
Perhaps this is the source of the seed field that is ultimately
responsible for the strong magnetic field in the core flares.

\par It is intriguing that in spite of the highly relativistically beamed Doppler core that the steep
spectrum secondary ($\alpha >1$) is mildly variable and has very
little apparent motion. This peculiar situation was studied in
section 6. A standard estimate of the long term time averaged
kinetic energy flux of the secondary reveals $\overline{Q}\approx
10^{42}\mathrm{ergs/sec}$ which is consistent with the core having
episodic flares of $Q \gtrsim 10^{43} \mathrm{ergs/sec}$, if the two
components are part of the same jetted beam of plasma. Thus, we
established consistency between the analysis of the core in section
5 and the kinetic energy flux of the secondary that was estimated in
section 6. We also noted that matched (to 5 GHz) resolution VLBA
observations indicates that the secondary is seen through a shroud
of free-free absorbing gas with an emission measure of $\approx
10^{8} \mathrm{cm}^{-6}\mathrm{pc}$. We argue that the steep
spectrum secondary seems to be a radio lobe associated with the jet
advancing into a dense medium (the jet is confined by ram pressure)
that is also the source of the free-free absorption. The properties
of the dense gas are consistent with a temperature, $10,000\,
\mathrm{K} \, - \, 20,000\,\mathrm{K}$, a displacement from the
nucleus, 3-4 pc, and a total hydrogen column density, $N_{H}=10^{22}
- 10 ^{23} \mathrm{cm}^{-2}$. These gas parameters are often
associated with the BAL wind \citep{ham03}. The implied kinetic
luminosity of this BAL wind is within the narrow range of values
that are predicted by the polar BAL wind theory in \citet{pun00}.
The density is at least an order of magnitude larger than what is
expected from such a wind, indicating that there is substantial gas
entrainment as the BAL wind propagates outward from the nucleus.

\par The jet in Mrk 231 might offer a unique opportunity to study the central
dynamics of a radio quiet quasar. Not only is the quasar near to
earth, but the jet is the brightest at high frequency of any radio
quiet quasar jet. Thus, no other radio quiet quasar central engine
can be explored with such high resolution (VLBI). Furthermore, now
we know the orientation of the jet. The high resolution of 43 GHz
VLBA allows us to peer into the inner regions of this quasar, using
the relativistic jet as a probe of the X-ray emitting gas, the X-ray
absorbing gas, and the BAL wind. In summary, there are many
potential discoveries that can be made with future VLBA monitoring
coordinated with observations at other frequency bands. Our
observations lead us to ask the the following questions
\begin {enumerate}
\item Is there more evidence of superluminal motion, like apparent
velocities of radio emission larger than the speed of light that can
be detected?
\item What is the maximum kinetic energy flux of the high frequency
flares?
\item Are these types of flares present in other radio quiet
quasars, but with varying magnitude?
\item Why are the flares so powerful, yet cool off so fast that MRK
231 never becomes a strong FR I radio source? More specifically, why
do all the components in the the core and secondary seem to show
such steep high frequency spectra? What is the powerful cooling
mechanism?
\item Is this cooling mechanism found in all radio quiet quasars and
why does it not occur in radio loud quasars?
\item Is the gas that entrains the secondary one and the same as the
BAL gas?
\item How fast do strong flares grow and what are the details of
their rapid decay?
\item Do the detected (but absorbed) X-rays come primarily from the same plasma
that is responsible for the 22 GHz flares (i.e., the accretion disk
X-rays are completely attenuated)? Do we see correlated flares in
the X-ray and microwave bands?
\item Since the microwave flare emission is from such small scales, it is
nearly coincident with the scales of X-ray emission and absorption.
Thus, can we expect a strong interplay between microwave flares and
the X-ray spectral properties of MRK231?
\item If the X-rays in Mrk 231 are predominantly inverse Compton emission from the microwave
jet, does the dense X-ray absorption column constrain the physical
parameters of the enveloping medium that gets entrained by the
propagating jet and thwarts it propagation to large scales?
\item Does a strong microwave flare blast holes in the X-ray
absorbing column allowing us to get a brief mildly attenuated
glimpse of the X-ray emission from the accretion disk corona?
\item Do flares in the microwave correspond to a decrease in the
X-ray emission of the accretion disk corona as has been hypothesized
as a universal phenomenon in black hole accretion systems (the
``low-hard" state)\citep{mac03,mar02}?
\end{enumerate}
Based on these considerations that were raised by our observations,
it seems clear that a coordinated observing program of VLBA and
X-ray observations would shed light on many of these issues, in
particular points 8 to 12. Generally speaking, coincident X-ray and
VLBA monitoring is potentially a powerful method of studying any
``disk-jet connection."

\end{document}